\documentclass[twoside, 12pt]{article}
\usepackage{pgfplots,comment,verbatim}
\usepackage[a4paper,top=1.18in, bottom=1.18in, left=1.2in, right=1.2in,bindingoffset=0mm]{geometry}

\usepackage{aditya}
\tikzset{
    >=latex,
    pil/.style={
            draw,
      <-, 
      decorate,
      decoration={snake,,amplitude=.02cm, pre length=.2cm,post length=.2cm,}
              }}

\hypersetup{colorlinks=true,linkcolor=red, citecolor= blue, urlcolor=darkred}

\definecolor{BluishGreen}{RGB}{0,158,115}

\title{{\bf \sc {{\color{darkblue}Skill Premia and Pre-Marital Investments in Marriage Markets
}}}\footnote{Kuvalekar: University of Essex and Indian School of Business, email: \texttt{a.kuvalekar@essex.ac.uk}. I am grateful to Nageeb Ali, Deepal Basak, Joyee Deb, Maitreesh Ghatak, Hugo Hopenhayn, Elliot Lipnowski, Dileep Mookherjee, Andrew Newman, Sabareesh Ramachandran, Debraj Ray, Anna Sanktjohanser, and participants in ThReD 2025 for helpful comments. Refine.ink was used to check the paper for consistency and clarity.}
}
\author{                   
\begin{minipage}{0.6\textwidth}\centering
Aditya Kuvalekar
\end{minipage}
}

\bigskip
\vspace{1cm}
\date{June 2026}

\begin{document}
\maketitle
\bigskip
\begin{center}
\end{center}
\bigskip
\begin{abstract}
I study a decentralized marriage market with search frictions, costly pre-marital skill investments, and non-transferable utility. Despite a fully symmetric environment, asymmetric equilibria---in which one gender systematically invests more in skills than the other---can arise. The match payoffs are microfounded through a non-cooperative household game in which spouses allocate time between labor-market work and domestic production. An asymmetric equilibrium becomes available precisely as the high-skill wage rises. Further, the symmetric equilibria can be fragile while the asymmetric ones are not. Thus, rising skill premia may amplify rather than narrow gender gaps in skill acquisition.

\bigskip

\vspace{0.2cm}

\end{abstract}
\newpage
\section{Introduction}
The wages commanded by highly skilled workers have been rising rapidly \citep{goldin2008race,acemoglu2011skills}. How are people's investments in skills affected by rising skill premia when their choice of skills has implications for both labor- and marriage-market outcomes? This is the central question of this paper.

From a labor-market standpoint, the implication seems clear: a higher return to skill should lead more people to incur the cost of skill acquisition. However, the answer is no longer so clear-cut when marriage-market considerations enter the picture.\footnote{That pre-marital investments respond to marriage-market returns, not only labor-market ones, is well documented: see \cite{lafortune2013} and \cite{chiappori2009investment}; \cite{goldin2006} documents how rising labor-market opportunities for women reshaped their educational investments over the twentieth century.} When the members of a household divide their time, one highly skilled person can focus on market work, while the other can focus on domestic production. Thus, an agent anticipating matching with a skilled partner may be discouraged from incurring the cost of investing. This within-household specialization can occur along gender lines, causing a widening gender gap in skills. 

In a canonical search-and-matching model with non-transferable utility, I show that gender disparities in skill acquisition can arise endogenously \emph{even when no primitive asymmetry exists between the genders}. 
These asymmetries arise precisely when the labor-market wage for high-skilled workers exceeds a threshold. Below the threshold, the equilibrium is symmetric: both genders invest at equal rates. Above the threshold, an asymmetric equilibrium emerges alongside the symmetric one. In any asymmetric equilibrium, one gender invests significantly more in skill acquisition than the other, who specializes in domestic production. Finally, in a class of markets, whenever an asymmetric equilibrium coexists with the symmetric one, the symmetric equilibrium is fragile in the sense of \cite{onuchic2023signaling} while the asymmetric one is robust to perturbations in investments.

The main contribution of this paper is to establish the existence and robustness of these asymmetric equilibria, and their emergence as the return to skill rises. In short, large shocks to the return to skill---of which AI is the most discussed contemporary example---need not affect men and women symmetrically, even when the shock itself is gender-neutral and the environment contains no structural discrimination. With marriage-market considerations at play, such a shock can widen the gap in skill acquisition between men and women rather than narrow it.

I now briefly describe the framework. A continuum of agents from two sides, men and women, engage in undirected search to find a potential match. Prior to beginning their search, each decides whether to incur a cost to acquire high skills or to remain low skilled. An agent's cost depends on their type, indexed by their identity in the unit interval; agents with higher types incur higher costs to acquire high skills. After making this investment choice, they begin the search phase, meeting partners from the other side randomly. When two people meet, they observe each other's skills and decide whether to match. Payoffs depend only on the two skills, not on latent types. If two agents match, they immediately leave the market and are replaced by newly born agents of the same types who make fresh investment decisions before entering the unmatched pool, ensuring that the economy is perpetually in steady state.

Before describing the results, I offer some interpretations of skill in the model. The most literal reading is education, the schooling an agent acquires before entering the marriage market. An alternative interpretation could be in terms of the career one pursues after basic education. The high-skill choice can represent a demanding, competitive, time-intensive career that pays well; the low-skill choice, a less onerous occupation with shorter hours and lower pay. 

I assume that everyone prefers a high-skilled partner to a low-skilled one. Therefore, negative assortative matching cannot arise, and pure-strategy equilibria fall into three classes: positive assortative matching (PAM), where high-skilled agents accept only high-skilled partners; all-match (AM), where everyone accepts everyone; and semi-assortative (SA), where high-skilled agents accept all partners but low-skilled agents reject low-skilled ones. What, then, could make two ex-ante identical sides of the market invest differently? Not PAM: any PAM equilibrium has both genders investing at the same rate (Proposition~\ref{Proposition: PAM is symmetric}, in the appendix). If asymmetry is to arise, it must come from the AM or SA classes.

Proposition~\ref{Proposition: equilibrium structure} characterizes what they allow. A symmetric equilibrium always exists, of AM or SA form. Alongside it the economy may also admit an \emph{asymmetric} equilibrium, in which one gender invests systematically more than the other; any such equilibrium takes one of two forms---\emph{full-investment-from-one-side} (FIOS), where one gender invests fully and the other only below a cutoff type, or \emph{no-investment-from-one-side} (NIOS), where one gender does not invest at all---and the two never coexist. Our focus is the asymmetric AM equilibria, which arise when search frictions are high enough that a low-skilled agent settles for a low-skilled partner rather than holding out for a high-skilled one.

These results show asymmetric equilibria are possible; a reader may still suspect they are a knife-edge curiosity. They are not. Whether one exists depends entirely on the match payoff function---the four payoffs for the possible own-and-partner skill combinations, and their relative magnitudes. This makes the question an empirical one: \cite{choo2006who}, \cite{galichon2022cupid}, and others have developed methods to identify match payoffs from observed matching patterns, so the conditions can, in principle, be checked in data. Proposition~\ref{Proposition: higher skill premia open asymmetric} then shows asymmetric equilibria are far from exceptional: from any economy whose only equilibrium is symmetric, raising the household's return to having a high-skilled member---the skill premium, read at the level of match payoffs---brings an asymmetric (FIOS) equilibrium into being alongside the symmetric one.

One question remains. Wherever an asymmetric equilibrium exists, a symmetric one exists beside it; is there any reason to expect the asymmetric outcome to be the one we see? Section~\ref{Section: fragility} answers through a recent refinement of \cite{onuchic2023signaling}, from their study of gender discrimination in collaborative projects, called \emph{fragility}. An equilibrium is fragile if a slight upward perturbation in one gender's investment, and downward in the other's, is amplified rather than undone by a single round of best-response play. Proposition~\ref{Proposition: fragility selection} shows that whenever an asymmetric AM equilibrium coexists with the symmetric one, it is the symmetric equilibrium that is fragile, while the asymmetric one is not. Far from pathological, the asymmetric equilibria are the robust ones.

The robustness of these asymmetric equilibria leads to the natural follow-up question: what changes in the underlying economy give rise to the changes in match payoffs that, in turn, induce asymmetric equilibria? Section~\ref{Section: microfoundation} provides one answer. I view the match payoff function as the equilibrium outcome of a non-cooperative household game along the lines of \cite{Chiappori1988}, in which spouses allocate time between labor-market work and household production.\footnote{See \cite{chiappori2020theory} for a survey of various models of household labor supply decisions as games.} Once match payoffs are themselves equilibrium objects rather than primitives, comparative statics can be performed on more fundamental objects of the economy---in particular, the labor-market wages paid to high- and low-skilled workers. The skill premium then has a concrete interpretation: the difference between the high-skill and low-skill wages. Proposition~\ref{Proposition: parametric comparative static} identifies a single threshold for the high-skill wage: below the threshold, where the payoffs are not yet submodular enough to admit an asymmetric equilibrium, the only equilibrium is symmetric, and the investment cutoff rises continuously with the premium as both genders invest more in skill acquisition; above the threshold, an asymmetric (FIOS) equilibrium becomes available---with one gender fully investing and the other investing substantially less---alongside the symmetric equilibrium.

In summary, the model delivers testable predictions. First, in the low-premium phase (below the threshold $t^*$), both genders invest at \emph{rising} and equal rates. Second, once the premium crosses $t^*$, an asymmetric equilibrium becomes available alongside the symmetric one: one gender invests considerably more in skill acquisition while the other specializes in domestic production. Under the fragility refinement of Section~\ref{Section: fragility}, the only robust outcome is the asymmetric equilibrium. Separately, at high premia the symmetric equilibrium can itself turn selective: low-skilled agents reject one another and match only with high-skilled partners, so low--low marriages disappear. Thus moderate increases in the skill premium raise both genders' investment equally, whereas a sufficiently large increase can lead one gender to lower its investment while the other raises its own significantly.

The rest of the paper is organized as follows. Section~\ref{Section: literature} discusses the related literature, and Section~\ref{Section: model} lays out the model. Section~\ref{Section: AM equilibria analysis} analyzes the post-investment matching game and the investment stage, establishing when asymmetric equilibria arise and how they respond to the skill premium. Section~\ref{Section: fragility} develops the fragility refinement and shows that the symmetric equilibrium is fragile while the asymmetric ones are not. Section~\ref{Section: microfoundation} microfounds the match payoffs through a household game and recasts the comparative statics in terms of labor-market wages. Section~\ref{Section: conclusion} concludes.

\section{Related Literature}\label{Section: literature}

Much of the extant literature on gender disparities in skill acquisition attributes them to primitive asymmetries between men and women. The present paper complements that line of work by showing that rising skill premia alone, in an otherwise symmetric environment, can generate the same kind of disparities.\footnote{I do not wish to suggest that structural asymmetries do not exist. My main focus, however, is on the comparative static---how pre-marital investments evolve with changing skill premia. It is \emph{a priori} not clear whether rising skill premia exacerbate these asymmetries or mitigate them. As Proposition~\ref{Proposition: parametric comparative static} shows, even in the absence of structural asymmetries, rising skill premia can give rise to asymmetries that would otherwise not be present.}

Naturally, this paper is related to the large literature on matching markets. There are two broad directions in which this literature has evolved. The initial papers, such as by \cite{becker1973theory,gale1962college,shapley1971assignment} assumed frictionless matching markets. Somewhat later, \cite{burdett1997marriage,shimer2000assortative,shimer2001matching,smith2006marriage} investigated these markets with search frictions. Across these two frameworks, another important distinction is between transferable utility (such as \cite{becker1973theory,shapley1971assignment}) vs non-transferable utility (such as \cite{gale1962college,smith2006marriage}). In frictionless NTU environments, \cite{legros2007beauty} characterize 
how the direction of assortative matching depends on the payoff structure 
in ways that diverge sharply from the TU case; the submodularity condition 
driving asymmetric equilibria in my paper mirrors their conditions for 
negative assortative matching. \cite{eeckhout1999} studies bilateral 
search with vertical heterogeneity under NTU, deriving sorting conditions 
that inform the post-investment matching stage of my model. The post-investment stage of my game can be viewed as a decentralized version of \cite{gale1962college} which was studied by \cite{adachi2003search,smith2006marriage}. To this setting, I add an investment choice before the search phase. 

A large literature has studied costly investments in marriage markets, \cite[see][for instance]{bhaskar2016marriage,bhaskar2023multidimensional,noldeke2015investment,cole2001efficient,peters2007pre,mailath2013pricing,chade2022risky}. Most of these papers study the efficiency of such investments. With the exception of \cite{chade2022risky} and \cite{bhaskar2016marriage} most other papers assume that the returns are deterministic, while these two study various issues related to how riskiness of investments affects the match outcomes as well as investments themselves. 
A seminal paper by \cite{peters2002competing} showed that, in a frictionless transferable-utility marriage market, non-cooperative premarital investments are nonetheless constrained efficient, as competition for partners provides the right investment incentives. I show that a combination of NTU and search frictions generates gender-specific asymmetries in such environments. Relatedly, \cite{bhaskar2023multidimensional} highlight how asymmetries in investments can occur when the two genders differ in their bargaining power. Also in a frictionless NTU environment, \cite{iyigun2007} show how
spousal allocations within the household feed back into premarital
investment incentives---a mechanism my paper embeds within a
search-and-matching framework. 

Recently, \cite{atakan2026efficient} study a marriage market with search frictions, costly investments, and \emph{transferable utility}. Their main finding is that the outcomes are constrained efficient and assortative in a wide variety of settings. In contrast, adopting the NTU framework, I sidestep the question of efficiency and focus solely on the disparities in skill acquisition that arise in equilibrium. 

Given my emphasis on gender differences in skill acquisition, this paper is naturally related to work that highlights these issues in various contexts, e.g.\ \cite{siow1998differntial,chiappori2009investment,zhang2021investment,chiappori2017partner,low2024human}. Often, unlike the current paper, these papers assume some innate differences---in preferences, fertility costs, or the technological environment---that give rise to these disparities. An exception is \cite{hadfield1999coordination}, who, like the present paper, derives gender-asymmetric specialization from symmetric primitives. Her framework, however, contains no labor market, and hence no analog of the comparative static in the skill premium that drives this paper. In a similar spirit, \cite{francois1998gender} is one of the early references to show how discrimination in the labor market can arise despite a fully symmetric environment, with the asymmetry emerging from the interaction between men and women within the household. \cite{lafortune2013} provides empirical support for the premise underlying this paper---that pre-marital investments respond to marriage-market returns to education, not only to labor-market returns---using variation in marriage-market competition across cohorts. On the
macroeconomic side, \cite{greenwood2014} document that rising assortative
mating has amplified income inequality, a consequence my mechanism
suggests can be driven partly by endogenous investment asymmetries rather
than match patterns alone.

\section{Model}\label{Section: model}

The economy consists of a unit mass of men and a unit mass of women, each indexed by a type $\type \in [0,1]$. We denote an agent's gender by $\g \in \{m, w\}$, and write $\gt$ for the opposite gender. Upon ``birth,'' agents choose a skill level from a finite set $\skillset$, which is assumed to be binary: $\skillset = \{H,L\}$, where $H$ represents high skill and $L$ represents low skill. The cost of acquiring skill $\skill$ for an agent of type $\type$ is $\cost(\type,\skill)$. Normalizing $\cost(\type,L) = 0$, we denote $\cost(\type,H)$ simply as $\cost(\type)$. Throughout the paper, we adopt an affine cost specification: $\cost(\type) = \type + c$ for some $c > 0$.\footnote{The possibility of asymmetric equilibria remains under general strictly increasing, strictly convex cost functions; the affine specification is adopted primarily for clean expressions that facilitate comparative statics.} Agents with higher $\type$ pay more to acquire $H$.

Although we label the two skill levels ``high'' and ``low,'' the relevant distinction need not be educational attainment as such. The high-skill option may equally be read as entry into a demanding, high-pressure, well-remunerated career track and the low-skill option as a less time-intensive, lower-paid one; under this reading an agent's investment decision is a choice of how ambitious a career to pursue, and the asymmetric equilibria below describe households in which one partner pursues such a career while the other does not.

If a man with skill $\skill$ and a woman with skill $\skill'$ match, the man’s payoff is $\payoff(\skill,\skill')$, and the woman’s is $\payoff(\skill',\skill)$. That is, the first argument represents an individual’s own skill and the second, their partner’s. This is a non-transferable utility (NTU) setting where payoffs depend only on skill levels, not on underlying type. Agents discount future payoffs at rate $r > 0$. 

We impose the following standard ranking of payoffs:
\bass\label{Assumption: payoff ranking}
$\phh\ge \phl \ge \payoff( L,H) \ge \pll$. \eass
This assumption can be microfounded through a household decision-making problem as in \cite{Chiappori1988}. See Section~\ref{Section: microfoundation} for an example.

I also restrict attention to equilibria in which every agent is matched eventually. Within the equilibrium classes we will consider---positive assortative matching (PAM), in which high types match only with high and low types only with low; all-match (AM), in which everyone accepts everyone; and semi-assortative (SA), in which high types accept all but low types reject low partners (all three are formally defined in Section~\ref{Section: equilibrium})---this rules out configurations in which some agents on one side of the market are systematically rejected by the other and remain permanently unmatched. This restriction is substantive rather than merely definitional: it sets aside coordination-driven equilibria in which one side is collectively unwilling to match with the other's low-skilled agents, leaving them permanently unmatched---outcomes sustained by self-fulfilling beliefs rather than by the investment incentives that are this paper's focus.

\noindent \textbf{An economy} is a tuple $\langle \payoff,\cost,\lm,r\rangle$ consisting of the payoff functions, the cost functions for acquiring different skill levels, and the arrival rate of potential partners that we will describe momentarily.

\noindent\textbf{Timing:}
Time is continuous and infinite. At each instant, unmatched agents engage in costless, undirected search, observing the skill distribution of the opposite pool. When two agents meet, they observe each other's skills and decide whether to match. If they do, they leave the market and are replaced by newly born agents of the same latent type $\type$ and the same gender as those who departed. (This assumption is without loss for our purposes; see Section~\ref{Section: discussion of model} for a discussion of alternative replacement rules.) These newly born agents make an irreversible skill investment decision before entering the unmatched pool.

Thus, the economy is perpetually in ``steady state'', i.e., the mass of each set of agents, men and women, remains to be $1$ and their type distribution remains unchanged. Unmatched agents meet others drawn at random from the unmatched pool at an exponential rate $\lm$.\footnote{Since the economy is perpetually in the steady state with a unit mass of population on either side, we do not need to scale it with the population size.} 

\noindent\textbf{Strategies:} We restrict attention to stationary equilibria (where strategies of any type do not vary with time). A strategy for a type $\type$ and gender $\g$ specifies a choice of skill to acquire, and then, subsequently, an acceptance set during the search phase. This specifies the set of agents one is willing to match with should they meet them. It is without loss to restrict attention to pure strategies insofar as investment is concerned. We denote by $\invest^\g: [0,1] \to \{0,1\}$ the investment strategy of a type $\type$ agent of gender $\g$, where one indicates their decision to acquire high skills and zero does not. We assume that $\invest^\g$ is measurable. Given players' investment strategies, let $\high_\invest^\g := \int \invest^\g(\type) \dd \type$ denote the proportion of $H$ types in group $\g$.  Since the dependence on $\invest$ is obvious in most cases, we will omit it. Post the agents' investment decisions, they engage in an undirected search. Whenever they meet someone, they have to decide whether to match with that person or not. If both the agents agree to a match, then the match takes place. The acceptance strategy specifies the probability of accepting an agent of each skill type and depends on one's own skill as well as the proportion of high types in the opposite pool. First, given Assumption~\ref{Assumption: payoff ranking}, no agent would ever turn down a high type agent. Therefore, we merely need to specify the probability of accepting a low type for each gender and skill type. We denote by $\accept^\g: [0,1]\times \{H,L\} \to [0,1]$ the probability that an agent of gender $\g$ and skill $\skill$ accepts a low type agent of the opposite gender given the proportion of the high types from group $\gt$. Often, this dependence on $\high^{\gt}$ is obvious and, therefore, we will suppress it.

\subsection{Discussion of the model}\label{Section: discussion of model}

The model has three key features: search frictions, costly pre-marital skill investments, and non-transferable utility. Search frictions and costly pre-marital investments are particularly pervasive in real marriage markets. Individuals---or their parents, in some contexts---make pre-marital investments anticipating labor-market outcomes that in turn shape their marriage prospects. Meanwhile, the prevalence of dating apps and marriage bureaus is clear evidence of search frictions in modern marriage markets.

The third feature, non-transferable utility (NTU), is motivated by the nature of social matches such as marriages. The seminal paper of \cite{becker1973theory} briefly considered such an environment, and \cite{smith2006marriage} studied a marriage market with search frictions and NTU at length. To quote the latter:
\begin{quote}
    \emph{But in defense of the NTU model, disagreements about matching in social settings are not uncommon. And whenever we observe a potential match or split desired by one party but not the other, utility is obviously not fully transferable. For the total match surplus either is positive or is not, and there can be no disagreement.}
\end{quote}
A particularly striking case of such a disagreement is the sharp discontinuity around $\tfrac{1}{2}$ in the share of household income earned by the wife documented by \cite{bertrand2015gender}. In the context of US couples, they show considerable aversion to situations in which the wife earns more than her husband. A world with fully transferable utility would show no such discontinuity: whatever psychological cost the husband incurs from his wife earning more could, in principle, be compensated through transfers.

Further, I have assumed that the flow payoff while unmatched is zero regardless of skill. This is largely a normalization. The qualitative results carry over to a setting in which an unmatched agent of skill $\skill$ earns a flow payoff $u(\skill) \ge 0$ with $u(H) \ge u(L)$, provided $u(\cdot)$ and the gap $u(H) - u(L)$ are sufficiently small relative to the match payoffs $\payoff(\cdot,\cdot)$. Intuitively, when the value of remaining unmatched is dwarfed by the value of marrying, the marriage market remains the dominant force shaping investment and acceptance decisions, and the structure of equilibria---including symmetry-breaking under submodularity and the comparative static on the skill premium---is preserved.

Let me briefly comment on the replacement rule. When two agents match and leave the market, they are replaced by newly born agents of the \emph{same latent type and same gender as those who just departed}: a departing type-$x$ man is replaced by a newly born type-$x$ man, and a departing type-$x$ woman by a newly born type-$x$ woman. The replacement agents then make fresh investment decisions before entering the unmatched pool. In any equilibrium, since agents of the same type and gender make identical investment choices, the replacements replicate the skill composition of the departing pair exactly. As a result, the type and skill composition of the unmatched pool is held fixed at the equilibrium investment shares throughout time, and the model remains in a stationary steady state without our having to track pool dynamics explicitly. This technology of replacement is adopted mainly for tractability: in every all-match (AM) equilibrium---one in which every agent accepts every partner they meet on the equilibrium path (formally defined in Section~\ref{Section: equilibrium})---every agent matches at rate $\lm$ regardless of skill, so the steady-state pool composition would coincide with the equilibrium investment share under any reasonable alternative replacement rule. The AM class covers most of the results in Sections~\ref{Section: AM equilibria analysis}--\ref{Section: microfoundation}.


\subsection{Equilibrium}\label{Section: equilibrium}
Agents choose their skill levels optimally given their expected value minus the cost. Let $\valfn^\g(\skill,\accept^\g;\high^{\gt},\accept^{\gt})$ denote the expected discounted value of an unmatched agent of gender $\g$ and skill $\skill$, given the high-skill share $\high^{\gt}$ in the opposite-gender pool and the acceptance strategies $\accept^\g,\accept^{\gt}$ of the two sides. These values solve standard steady-state recursive equations; the expressions are straightforward to compute, so we omit them. An agent of skill $\skill$ and gender $\g$ chooses $\accept^\g$ to maximize $\valfn^\g(\skill,\accept^\g;\high^{\gt},\accept^{\gt})$ given $(\high^{\gt},\accept^{\gt})$.

\noindent \textbf{An equilibrium} consists of the investment and acceptance strategies for each gender and skill type, $\langle (\invest^\g, \accept^\g)_{\g \in \{m,w\}}\rangle$, satisfying \eqref{Equation: IC to invest} and \eqref{Equation: IC to accept low types} below.

\begin{align}
    \valfn^\g(H,\accept^\g;\high^{\gt},\accept^{\gt}) - \cost(\type) - \valfn^\g(L,\accept^\g;\high^{\gt},\accept^{\gt})\ & \begin{cases} > 0 \implies \invest^\g(\type) = 1 \\
    =0 \implies \invest^\g(\type) \in \{0,1\} \\
    < 0 \implies \invest^\g(\type) = 0 \end{cases} \tag{IC-invest} \label{Equation: IC to invest}\\
    \valfn^\g(\skill,\accept^\g;\high^{\gt},\accept^{\gt})\  & \begin{cases} < \payoff(\skill,L) \implies \a^\g(\skill) = 1 \\
    = \payoff(\skill, L) \implies \a^\g(\skill) \in [0,1]\\
    > \payoff(\skill,L) \implies \a^\g(\skill) = 0 \end{cases} \tag{IC-AL}\label{Equation: IC to accept low types}
\end{align}
\eqref{Equation: IC to invest} ensures that the agents' investment decisions are optimal, while \eqref{Equation: IC to accept low types} specifies that agents acceptance decisions are optimal. In particular, if the value to agent is higher than what (s)he would receive by matching with a low-skilled person, then (s)he must not be accepting the low-skilled person. Because the value functions do not depend on an agent's own type while the cost $\cost(\type) = \type + c$ is strictly increasing in it, the net gain to investing is strictly decreasing in $\type$; optimal investment is therefore a cutoff rule. In any equilibrium, the investment strategies are characterized by a cutoff type, $\ttypecg$ for each $\g$, such that an agent $i$ of gender $\g$ and type $\type_i$ invests if and only if $\type_i \le \ttypecg$.

\bnot In light of the above, we will denote an equilibrium by $(\ttype,\accept)$ where each, $\ttype= (\ttypem,\ttypew)$ and $\accept = (\accept^\g(H),\accept^\g(L))_{\g \in \{m,w\}}$ are vectors denoting the relevant strategies. In general, $\typev$ denotes a vector $(\type_m,\type_w)$. \enot

We turn now to the structure of sorting in equilibrium and, more importantly, the investments agents make. In light of Assumption~\ref{Assumption: payoff ranking}, no agent will ever reject a high skilled person. Therefore, we can rule out the existence of a negatively assortative match.  Thus, insofar as pure strategy equilibria are concerned, we are left with the following three possibilities:
\begin{enumerate}
    \item All match (AM): All agents accept all partners.
    \item Positive Assortative Matching (PAM): High types match only with high types, and low types only with low types.
    \item Semi-assortative matching (SA): High types accept all partners, but low types reject low-skill partners and match only with high types.
\end{enumerate}
A fourth possible candidate---in which one side of the market rejects all low-skill partners from the other side, leaving those low-skill agents permanently unmatched---is excluded by the restriction to equilibria in which every agent is matched eventually.

The bulk of our analysis focuses on AM and SA equilibria; PAM is analyzed separately in Appendix~\ref{Section: PAM and SA appendix}. Both classes of symmetric equilibrium (AM and SA) play roles in the analysis below: which one obtains depends on whether low-skilled agents prefer to accept or reject low-skilled partners at the equilibrium cutoff. PAM, by contrast, requires the marriage market to be sufficiently thick---high types must be willing to wait through encounters with low-type partners for a high-type match---and its detailed analysis is deferred to the appendix.

\section{Analysis}\label{Section: AM equilibria analysis}

The analysis throughout assumes submodular payoffs ($\De \le 0$). We focus on this case because under supermodular payoffs ($\De \ge 0$), all equilibria are symmetric---asymmetric investment outcomes cannot arise (see Proposition~\ref{Proposition: unique symmetric AM eqm if SM} in the appendix). The economically interesting question of when asymmetric equilibria can emerge from a symmetric environment therefore arises precisely in the submodular regime. Recall from Section~\ref{Section: equilibrium} that pure-strategy equilibria fall into three classes: AM (all accept all), PAM (high accept only high, so low match only with low), and SA (high accept all, low reject low). PAM is discussed in Appendix~\ref{Section: PAM analysis}; here we analyze the AM and SA classes that drive the paper's substantive results.

\subsection*{AM equilibria}

In an AM equilibrium with cutoffs $\ttype = (\ttypem, \ttypew)$, every agent accepts every partner, so $\a^\g(\cdot) = 1$ for all $\g$. The players' values are:
\begin{align*}
    \valfn_A^\g(H; \ttype) =& \frac{\lm [\ttypec_{\gt} \phh+ ( 1- \ttypec_{\gt})\phl]}{r + \lm}, \quad \text{ and } \\
    \valfn_A^\g(L; \ttype) =& \frac{\lm [\ttypec_{\gt} \plh+ ( 1- \ttypec_{\gt})\pll]}{r + \lm}.
\end{align*}
Ignoring trivial cases where either everybody invests or nobody does, and assuming at least one of $\ttypem, \ttypew$ is interior, the IC constraints are:
\begin{align}
    \valfn^\g_A(H; \ttype) - \cost(\ttypecg) =&\ \valfn^\g_A(L; \ttype) \quad \text{ if } \ttypecg \in (0,1) \tag{AM: Invest} \label{Equation: AM invest IC}\\
    \valfn^\g_A(H; \ttype) \ \le&\ \phl \tag{AM: H-L}\label{Equation: AM H accepts L}\\
    \valfn^\g_A(L; \ttype) \ \le&\ \pll \tag{AM: L-L}\label{Equation: AM L accepts L}
\end{align}
Here, \eqref{Equation: AM invest IC} says the cutoff type $\ttypecg$ is indifferent between investing and not. \eqref{Equation: AM H accepts L} says that a high-skilled agent's continuation value of remaining in the search pool does not exceed the payoff $\phl$ from accepting a low-skilled partner---so a high-skilled agent prefers to match with a low-skilled partner rather than wait. \eqref{Equation: AM L accepts L} is the analogous statement for a low-skilled agent: their continuation value does not exceed $\pll$, so they too accept a low-skilled partner on meeting.

\subsection*{SA equilibria}

In a (symmetric) SA equilibrium with common cutoff $\widehat{x} \in (0, 1)$, low-type agents reject low-type partners while high-type agents accept all. The pool on each side has share $\widehat{x}$ in $H$. Match rates differ by skill: H matches at rate $\lm$, L matches at rate $\lm \widehat{x}$ (only with H). The values are:
\begin{align*}
    \valfn^\g_S(H; \widehat{x}) &= \frac{\lm[\widehat{x}\, \phh + (1-\widehat{x})\, \phl]}{r + \lm} \quad \text{(same as AM)} \\
    \valfn^\g_S(L; \widehat{x}) &= \frac{\lm \widehat{x}}{r + \lm \widehat{x}}\, \plh
\end{align*}
For later reference, denote the value-of-investing gap in the SA regime at common pool composition $\widehat{x}$ by
\begin{equation}\label{Equation: G definition}
G(\widehat{x}) \,:=\, \valfn^\g_S(H; \widehat{x}) - \valfn^\g_S(L; \widehat{x}) \,=\, \frac{\lm[\widehat{x}\,\phh + (1-\widehat{x})\,\phl]}{r+\lm} - \frac{\lm \widehat{x}\,\plh}{r + \lm \widehat{x}}.
\end{equation}

The IC constraints are:
\begin{align}
    \valfn^\g_S(H; \widehat{x}) - \cost(\widehat{x}) =&\ \valfn^\g_S(L; \widehat{x}) \tag{SA: Invest}\label{Equation: SA invest IC}\\
    \valfn^\g_S(L; \widehat{x}) >&\ \pll \tag{SA: L-L Reject}\label{Equation: SA L-L reject}
\end{align}
The L-rejection IC \eqref{Equation: SA L-L reject} simplifies to $\lm \widehat{x}(\plh - \pll) > r\,\pll$. It is the exact negation of the AM L-L acceptance constraint at the same cutoff: at any given $\widehat{x}$, only one acceptance regime is optimal for the L type.

\subsection*{Closed-form cutoff candidates}

It is convenient to summarize the payoff parameters that drive the analysis. Define
\begin{align*}
    \De &:= \phh + \pll - \phl - \plh \quad (\le 0 \text{ under submodularity}), \\
    \De_h &:= \phl - \pll \quad (\ge 0), \\
    \eta &:= \plh - \pll \quad (\ge 0).
\end{align*}
By Assumption~\ref{Assumption: payoff ranking}, $\De_h \ge \eta \ge 0$. Throughout the paper, ``SYM AM'' refers to a symmetric AM equilibrium (both genders at the same cutoff under all-match acceptance), and ``SYM SA'' to a symmetric SA equilibrium (both genders at the same cutoff with high types accepting all and low types rejecting low partners).

The closed-form cutoff candidates for the SYM AM, FIOS, and NIOS equilibria are:
\begin{equation}
\begin{aligned}
    \ttsym &:= \max\!\left\{0,\, \min\!\left\{1,\, \frac{\lmr \De_h - c}{1 - \lmr \De}\right\}\right\} && \text{(SYM AM)}, \\
    \ttlb  &:= \max\!\left\{0,\, \min\!\left\{1,\, \lmr (\De + \De_h) - c\right\}\right\} && \text{(FIOS)}, \\
    \ttub  &:= \max\!\left\{0,\, \min\!\left\{1,\, \lmr \De_h - c\right\}\right\} && \text{(NIOS)}.
\end{aligned}
\label{Equation: cutoff candidates}
\end{equation}

By construction, each candidate lies in $[0,1]$; the formulae in parentheses in \eqref{Equation: cutoff candidates} coincide with the cutoffs whenever they are interior, and the boundary values correspond to corner equilibria where one skill type is absent. Under submodularity ($\De \le 0$), $\ttlb \le \ttsym \le \ttub$. We define the FIOS profile as $(1, \ttlb)$ (all of one side invest; the other side has cutoff $\ttlb$) and the NIOS profile as $(0, \ttub)$ (no investment on one side; the other has cutoff $\ttub$); both are described up to permutation of genders. In a FIOS profile the fully-investing side has no low-skilled agents on the equilibrium path, so matching is all-match on path; the low-skilled acceptance of that side is an off-path object, pinned by the optimal behavior of an agent who deviates to low skill.\footnote{This is what disciplines the fully-investing side's incentive to invest. An agent of the fully-investing gender who deviated to low skill would face an opposite pool of high-skill share $\ttlb$ and would optimally \emph{reject} low partners whenever $\lm\ttlb\,\eta > r\pll$, securing the value $\valfn_S(L;\ttlb) = \lm\ttlb\,\plh/(r+\lm\ttlb)$ rather than the all-match value $\valfn_A(L;\ttlb)$. The FIOS profile is therefore an equilibrium only if its cutoff type weakly prefers to invest against this \emph{optimal} deviation value, $\valfn_A(H;\ttlb) - (1+c) \ge \max\{\valfn_A(L;\ttlb),\,\valfn_S(L;\ttlb)\}$, equivalently
\[
    \min\bigl\{\lmr(\ttlb\De+\De_h),\ G(\ttlb)\bigr\} \ \ge\ 1+c .
\]
When $\lm\ttlb\,\eta\le r\pll$ this reduces to the all-match condition $\lmr(\ttlb\De+\De_h)\ge 1+c$ (FIOS is genuinely all-match); when $\lm\ttlb\,\eta> r\pll$ it is the strictly stronger requirement $G(\ttlb)\ge 1+c$. NIOS raises no such subtlety: there the low-skilled side is populated on path, so its L-L acceptance constraint is imposed directly (see the proof of Proposition~\ref{Proposition: equilibrium structure}).} The SYM SA cutoff $\ttsa$, when it arises, is defined implicitly by the SA investment IC \eqref{Equation: SA invest IC} and is distinct from $\ttsym$.

\subsection*{Equilibrium structure}

Throughout the AM/SA analysis we maintain the condition
\begin{equation}
    \lm\,(\phh - \phl) < r\, \phl, \tag{$\star$}\label{Equation: no PAM}
\end{equation}
which says that a high-skilled agent always strictly prefers to accept a low-skilled partner rather than wait. This is precisely what it means to focus on AM and SA equilibria: \eqref{Equation: no PAM} rules out PAM (Proposition~\ref{Proposition: PAM is symmetric}) and ensures the symmetric equilibrium takes AM or SA form.

Before we see the formal results, I offer a quick preview of them here. Proposition~\ref{Proposition: equilibrium structure} characterizes the equilibrium set: a symmetric equilibrium always exists, all-match (SYM AM) or semi-assortative (SYM SA), and alongside it the economy may admit asymmetric equilibria, in which one gender invests more than the other, of the FIOS or NIOS form. Proposition~\ref{Proposition: higher skill premia open asymmetric} shows these are not a curiosity: starting from any economy whose only equilibrium is SYM AM, raising the skill premium constructs an economy that admits an asymmetric equilibrium. Proposition~\ref{Proposition: rising investment} records a comparative static: as the premium rises, the investment cutoff rises with it (in the symmetric equilibrium), so more agents acquire skill. Finally, Proposition~\ref{Proposition: fragility selection} shows that whenever an asymmetric equilibrium coexists with the symmetric one, the symmetric equilibrium is fragile while the asymmetric one is not, so the asymmetric outcomes, once available, are the robust ones. Section~\ref{Section: microfoundation} then grounds the skill premium in labor-market wages through a household game.

\bprop[Equilibrium structure]\label{Proposition: equilibrium structure}
Maintain \eqref{Equation: no PAM}.
\begin{enumerate}
    \item \emph{Symmetric equilibrium.} A symmetric equilibrium always exists and is unique. It is a SYM AM equilibrium at cutoff $\ttsym$ when low-skilled agents accept low-skilled partners there ($\lm \ttsym\, \eta \le r\, \pll$); otherwise SYM AM does not exist and the symmetric equilibrium is SYM SA, at the unique cutoff $\ttsa < \ttsym$.
    \item \emph{Configurations with SYM AM.} Suppose SYM AM exists, and set aside the knife-edge $1 + \lmr \De = 0$.\footnote{At the knife-edge $1 + \lmr \De = 0$---a non-generic, measure-zero locus in the payoff parameters---the two interior investment indifference conditions become collinear, and a continuum of interior asymmetric profiles with $\ttypem + \ttypew = 2\ttsym$ can be sustained. We set this case aside throughout.} Any asymmetric (AM) equilibrium is then either FIOS $(1, \ttlb)$ or NIOS $(0, \ttub)$, up to permutation of genders, and the two never coexist. Consequently the economy realizes exactly one of three configurations: SYM AM is the unique equilibrium; SYM AM coexists with a NIOS equilibrium and its mirror image; or SYM AM coexists with a FIOS equilibrium and its mirror image.
\end{enumerate}
\eprop

Figure~\ref{Figure: equilibrium structure} illustrates the configurations of Proposition~\ref{Proposition: equilibrium structure}. The proposition lays out the possible equilibrium configurations but does not say which one a given environment realizes---in particular, whether an asymmetric equilibrium is available at all. That is the role of Proposition~\ref{Proposition: higher skill premia open asymmetric} below, which shows that asymmetric (FIOS) equilibria emerge as the skill premium rises.

\begin{remark}\label{Remark: not exhaustive}
Two configurations fall outside Proposition~\ref{Proposition: equilibrium structure}. PAM arises when \eqref{Equation: no PAM} fails, and is treated in Appendix~\ref{Section: PAM and SA appendix}. And when SYM AM does not exist---so the symmetric equilibrium is SYM SA---asymmetric equilibria in which both genders are selective, at different cutoffs, may also arise; we do not pursue these, since our interest is in the asymmetric AM equilibria that emerge as the skill premium rises (Proposition~\ref{Proposition: higher skill premia open asymmetric}).
\end{remark}

\begin{remark}[Welfare]
When a symmetric and an asymmetric equilibrium coexist, they are Pareto incomparable: the fully-investing side strictly prefers the symmetric outcome, while the other side strictly prefers the asymmetric one. We omit the formal argument, which amounts to comparing the partner pool each side faces across the two equilibria.
\end{remark}

\begin{center}
\begin{minipage}{\textwidth}
\centering
\begin{tikzpicture}[x=1cm,y=0.8cm,
   inv/.style={line width=3.4pt, color=darkblue, line cap=round},
   non/.style={line width=3.4pt, color=red, line cap=round},
   tk/.style={line width=0.9pt, color=black!85},
   lab/.style={font=\footnotesize},
   grp/.style={font=\bfseries\footnotesize, anchor=west}]

\def\W{7.4}
\def\xlb{1.1}    
\def\xA{3.3}     
\def\xSA{2.6}    
\def\xub{5.5}    

\draw[inv] (0,9.3) -- (0.7,9.3);
\node[lab,anchor=west] at (0.85,9.3) {acquire $H$ (invest)};
\draw[non] (4.0,9.3) -- (4.7,9.3);
\node[lab,anchor=west] at (4.85,9.3) {remain $L$ (do not invest)};

\node[grp] at (-4.6,8.25) {Symmetric AM};
\node[lab,anchor=east] at (-0.2,8.5) {Men};
\draw[non] (0,8.5) -- (\W,8.5);
\draw[inv] (0,8.5) -- (\xA,8.5);
\draw[tk] (\xA,8.35) -- (\xA,8.65);
\node[lab,anchor=east] at (-0.2,8.0) {Women};
\draw[non] (0,8.0) -- (\W,8.0);
\draw[inv] (0,8.0) -- (\xA,8.0);
\draw[tk] (\xA,7.85) -- (\xA,8.15);
\node[lab,anchor=north] at (\xA,7.8) {$\ttsym$};
\node[lab,anchor=west,color=black!55] at (\W+0.15,8.25) {$L$ accepts $L$};

\node[grp] at (-4.6,6.65) {Symmetric SA};
\node[lab,anchor=east] at (-0.2,6.9) {Men};
\draw[non] (0,6.9) -- (\W,6.9);
\draw[inv] (0,6.9) -- (\xSA,6.9);
\draw[tk] (\xSA,6.75) -- (\xSA,7.05);
\node[lab,anchor=east] at (-0.2,6.4) {Women};
\draw[non] (0,6.4) -- (\W,6.4);
\draw[inv] (0,6.4) -- (\xSA,6.4);
\draw[tk] (\xSA,6.25) -- (\xSA,6.55);
\node[lab,anchor=north] at (\xSA,6.2) {$\ttsa$};
\node[lab,anchor=west,color=black!55] at (\W+0.15,6.65) {$L$ rejects $L$};

\node[grp] at (-4.6,5.05) {Asym.\ FIOS};
\node[lab,anchor=east] at (-0.2,5.3) {Men};
\draw[inv] (0,5.3) -- (\W,5.3);
\node[lab,anchor=west,color=black!55] at (\W+0.15,5.3) {all invest};
\node[lab,anchor=east] at (-0.2,4.8) {Women};
\draw[non] (0,4.8) -- (\W,4.8);
\draw[inv] (0,4.8) -- (\xlb,4.8);
\draw[tk] (\xlb,4.65) -- (\xlb,4.95);
\node[lab,anchor=north] at (\xlb,4.6) {$\ttlb$};

\node[grp] at (-4.6,3.45) {Asym.\ NIOS};
\node[lab,anchor=east] at (-0.2,3.7) {Men};
\draw[non] (0,3.7) -- (\W,3.7);
\node[lab,anchor=west,color=black!55] at (\W+0.15,3.7) {none invest};
\node[lab,anchor=east] at (-0.2,3.2) {Women};
\draw[non] (0,3.2) -- (\W,3.2);
\draw[inv] (0,3.2) -- (\xub,3.2);
\draw[tk] (\xub,3.05) -- (\xub,3.35);
\node[lab,anchor=north] at (\xub,3.0) {$\ttub$};
\end{tikzpicture}
\captionsetup{font=footnotesize}

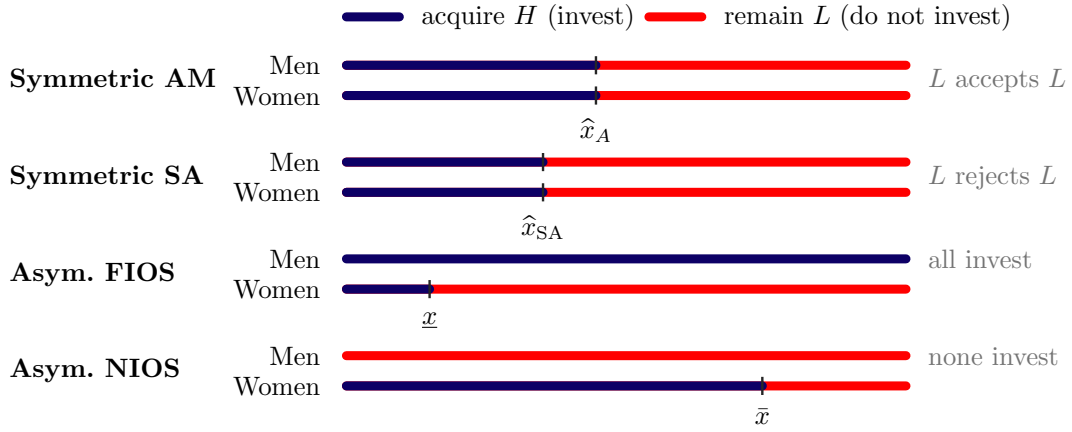
\captionof{figure}{The equilibrium configurations of Proposition~\ref{Proposition: equilibrium structure}. Each horizontal line is the type space $[0,1]$ for one gender; the blue segment marks the types that acquire high skill, the red segment those that remain low-skilled. The two symmetric equilibria share a common cutoff across genders and differ in acceptance behavior: under \emph{Symmetric AM} low-skilled agents accept low-skilled partners (cutoff $\ttsym$), while under \emph{Symmetric SA} they reject them (cutoff $\ttsa < \ttsym$). In \emph{FIOS} one gender invests fully while the other invests only below $\ttlb$; in \emph{NIOS} one gender does not invest at all while the other invests below $\ttub$. The cutoffs satisfy $\ttlb \le \ttsym \le \ttub$. Gender labels are illustrative: each asymmetric configuration has a mirror image with the roles of men and women swapped.}
\label{Figure: equilibrium structure}
\end{minipage}
\end{center}

\medskip

Proposition~\ref{Proposition: equilibrium structure} establishes the conditions under which an asymmetric equilibrium can surface alongside the symmetric one. These conditions are stated in terms of the primitives of the surplus functions. Next, we present a simple comparative static of how changes in the underlying surplus functions can give rise to such asymmetric equilibria. \cite{choo2006who}, \cite{galichon2022cupid}, and others have developed methods to estimate marriage surplus (match payoff) functions from observed matching patterns, which makes a comparative static phrased directly in terms of these surplus functions empirically meaningful. Proposition~\ref{Proposition: higher skill premia open asymmetric} establishes the pervasiveness of asymmetric equilibria as the skill premium---the increase in each household member's payoff from having at least one highly-skilled member---rises.

\bdefn\label{Definition: skill premium}We say that an economy $\bpayoffh := \langle \phhhat, \phlhat, \plhhat, \pllhat\rangle$ exhibits a \emph{higher skill premium} relative to $\bpayoff := \langle \phh, \phl, \plh, \pll\rangle$ if $\phhhat \ge \phh, \phlhat \ge \phl, \plhhat \ge \plh$ and $\pllhat = \pll$. \edefn

\bprop[Higher skill premia make asymmetric equilibria possible]\label{Proposition: higher skill premia open asymmetric}
Fix $r, \lm, c$. Suppose that the unique symmetric equilibrium of an economy $\bpayoff$ with $\De \le 0$ is SYM AM.\footnote{A simple sufficient condition for the symmetric equilibrium to be all-match is $\lm\eta \le r\pll$: a low-skilled agent always prefers to marry a low-skilled partner she meets over holding out for a high-skilled one. This essentially requires the search frictions to be sufficiently large---the meeting rate $\lm$ small relative to the discount rate $r$---so that waiting for a better-skilled partner is not worthwhile, however valuable such a partner becomes as the premium rises. We do not impose this condition formally.} Furthermore, suppose that
\begin{enumerate}[(i)]
    \item $1 + \lmr \De > 0$, and
    \item $\lmr \De_h - c \in (0, 1)$.\footnote{Condition (ii) is not essential to the proof but simplifies the casework. It ensures the symmetric cutoff $\ttsym$ is interior.}
\end{enumerate}
Then there exists an economy $\bpayoffh$ exhibiting a higher skill premium relative to $\bpayoff$ such that $\bpayoffh$ admits an asymmetric (FIOS) equilibrium.
\eprop

To illustrate, the following numerical example walks through the construction.

\bexam\label{Example: comparative static}Suppose that $c = 2$, $r = \lm = 1$, $\pll = 1$. Consider the following two environments.
\begin{enumerate}[(i)]
    \item $\phh = 7, \phl = 6, \plh = 3$.
    \item $\phhhat = 11, \phlhat = 10, \plhhat = 6$.
\end{enumerate}
\eexam

In environment $(i)$, the only equilibrium is symmetric with cutoff $\ttsym = \tfrac{1}{3}$. In environment $(ii)$, the symmetric cutoff $\ttsym = \tfrac{5}{6}$ no longer constitutes an equilibrium (the SYM AM L-L acceptance fails), so the symmetric equilibrium is instead a SYM SA at a slightly different cutoff; in parallel, an asymmetric FIOS equilibrium emerges at $\ttlb = 0.5$. Here $\lm\ttlb\,\eta = 0.5\cdot 5 = 2.5 > r\pll = 1$, so a fully-investing agent who deviated to low skill would reject low partners; the investing-side IC is accordingly tested against the deviation-optimal value $\valfn_S(L;\ttlb)=\lm\ttlb\,\plh/(r+\lm\ttlb)=2$, and $\valfn_A(H;\ttlb)-(1+c)=5.25-3=2.25\ge 2$ confirms the FIOS profile is an equilibrium (all-match on path). The difference between the two environments lies entirely in the payoffs where at least one partner is highly skilled, which I interpret as a rise in the skill premium.

Proposition~\ref{Proposition: higher skill premia open asymmetric} shows that asymmetric equilibria become available as the skill premium rises. Before turning to how an economy selects among the equilibria available to it, we record what happens to the symmetric equilibrium itself as the premium rises. Index a family of economies by a parameter $\tau$, a higher value of which corresponds to a higher skill premium; concretely, $\tau$ can be taken to be the labor-market wages for high- and low-skilled workers, through which Section~\ref{Section: microfoundation} microfounds the match payoffs $\phh, \phl, \plh$.

\bprop[Investment rises with the premium]\label{Proposition: rising investment}
Let $\phh(\tau)$, $\phl(\tau)$, $\plh(\tau)$ be continuously differentiable and non-decreasing in $\tau$, with $\pll$ fixed, and suppose the symmetric equilibrium is SYM AM with interior cutoff $\ttsym(\tau)$. If $\De_h'(\tau) \ge |\De'(\tau)|$, then $\ttsym(\tau)$ is non-decreasing in $\tau$, and strictly increasing wherever the inequality is strict.
\eprop

Observationally, then, we would find more people investing in skills as the skill premium rises, provided these private returns rise faster than the submodularity. Beyond a certain threshold, however, an asymmetric equilibrium emerges (Proposition~\ref{Proposition: higher skill premia open asymmetric}). If society coordinates on it, we would instead see a divergence in the genders' investments: one gender invests more, the other less. This raises a natural question: why would society switch to an asymmetric equilibrium when a symmetric one remains available? The next section shows that whenever the asymmetric equilibria become available, the symmetric all-match equilibrium becomes fragile in the sense of \cite{onuchic2023signaling}---in a sense, the asymmetric equilibria are robust to perturbations in investment that their symmetric counterpart is not. We turn to this next.

\section{Equilibrium Selection}\label{Section: fragility}

Proposition~\ref{Proposition: equilibrium structure} establishes the possibility of an asymmetric equilibrium. However, whenever an asymmetric equilibrium exists, it is accompanied by a symmetric one. We now evaluate the robustness of these equilibria and argue that the symmetric equilibria are \emph{fragile}, where we adopt the notion of fragility due to \cite{onuchic2023signaling}. Before presenting the notion formally, we briefly describe it in words.

Roughly, an equilibrium is fragile if a small antisymmetric perturbation---one gender's investment cutoff nudged up, the other's nudged down by the same amount---gets amplified by a single round of best-response play, in the sense that both genders' cutoffs move further from the original equilibrium than the perturbation itself. The intuition is that even a small payoff-irrelevant bias across gender lines can compound into a sizable deviation, undermining the equilibrium as a focal point. Equilibria that pass this test, by contrast, absorb such perturbations rather than amplifying them.

\subsection*{The notion of fragility}

The joint best-response iteration on the cutoff vector $(\ttypem, \ttypew) \in [0,1]^2$ depends on the acceptance regime of the equilibrium under consideration. In the AM regime, define the investment best-response
\[
\mathrm{BR}_{AM}(y) \,:=\, \max\!\big\{0,\, \min\!\big\{1,\, \lmr(\De\,y + \De_h) - c\big\}\big\},
\]
and in the SA regime,
\[
\mathrm{BR}_{SA}(y) \,:=\, \max\!\big\{0,\, \min\!\big\{1,\, G(y) - c\big\}\big\}.
\]
Writing $\mathrm{BR}$ for whichever applies, the joint iteration is
\[
\Theta(\ttypem, \ttypew) \,:=\, \bigl(\mathrm{BR}(\ttypew),\, \mathrm{BR}(\ttypem)\bigr),
\]
and equilibria of the corresponding class are the fixed points of $\Theta$. We adapt \citet{onuchic2023signaling}'s definition of fragility to our setting:

\bdefn[Fragility; \citealp{onuchic2023signaling}]\label{Definition: fragility}
An equilibrium $(\ttypem^*, \ttypew^*)$ of $\Theta$ is \emph{fragile} if there exist $\delta > 0$ and $\zeta > 0$ such that for every $\epsilon \in (-\delta, \delta)$ for which the perturbation $(\ttypem^* + \epsilon, \ttypew^* - \epsilon)$ lies in $[0,1]^2$,
\begin{align*}
\bigl|\,\Theta_m(\ttypem^* + \epsilon, \ttypew^* - \epsilon) - \ttypem^*\,\bigr| &\ge |\epsilon|(1+\zeta) \quad\text{and}\\
\bigl|\,\Theta_w(\ttypem^* + \epsilon, \ttypew^* - \epsilon) - \ttypew^*\,\bigr| &\ge |\epsilon|(1+\zeta),
\end{align*}
where the subscripts on $\Theta$ refer to its component functions.
\edefn

The feasibility qualifier $(\ttypem^* + \epsilon, \ttypew^* - \epsilon) \in [0,1]^2$ is needed because cutoffs are bounded; at a corner equilibrium only one sign of $\epsilon$ remains feasible, and the condition is tested on that portion of $(-\delta, \delta)$.

\bprop[Fragility selection]\label{Proposition: fragility selection}
Maintain \eqref{Equation: no PAM} and the affine cost specification. Suppose SYM AM and an asymmetric (AM) equilibrium---FIOS or NIOS, with interior cutoff---coexist. Then SYM AM is fragile, while the coexisting asymmetric equilibrium is not.
\eprop

Thus, whenever a symmetric AM equilibrium coexists with an asymmetric (FIOS or NIOS) equilibrium, the symmetric one is fragile while the asymmetric one is not. Consequently, for a non-knife-edge set of parameters, the only equilibria that survive the fragility test are the asymmetric ones; since the environment is fully symmetric, these arise as mirror-image pairs, between which the refinement does not select. Section~\ref{Section: microfoundation} expresses the underlying threshold in terms of observable labor-market wages.

\section{Microfoundation from household maximization in NTU}\label{Section: microfoundation}
The preceding sections established that asymmetric equilibria can arise---under conditions stated entirely in terms of the match payoff function---whenever the payoff function is sufficiently submodular. This raises a natural question: what changes in the underlying economy give rise to the changes in match payoffs that, in turn, induce asymmetric equilibria? This section provides one answer. I view the match payoff function as the equilibrium outcome of a non-cooperative household game along the lines of \cite{Chiappori1988},\footnote{\citet{Chiappori1988} adopts a cooperative formulation to obtain the Pareto-optimal choices. While I keep the preferences of the members of the household similar, I adopt a non-cooperative approach in determining the labor supply choices. See \citet{lundberg1993separate} for similar models.} in which a household consisting of a man and a woman allocates two units of time between labor-market work and household production. The match payoffs $\phh, \phl, \plh, \pll$ thus become equilibrium objects of the household game, with the labor-market wages of the two members serving as fundamental primitives. The setup serves two purposes. First, it shows that match payoff functions supporting asymmetric equilibria can arise endogenously from plausible labor-market and household primitives. Second, it allows us to perform comparative statics on more fundamental objects than the surplus function itself---most notably, the wage structure---and to ask how rising wages for high-skilled workers translate into changes in marriage-market behavior, producing what one might call ``skill inequality'' between genders.

Let us work backwards to see the idea clearly. The households' problem is the following. A household consists of a man $m$ and a woman $w$ with skills $s(m)$ and $s(w)$. Wages in the market depend only on the skill and not on the gender.\footnote{As would be clear, a gender pay gap as it exists in reality can only make it easier to construct equilibria where women invest less in skill acquisition.} The households decide how to split their two units of time---one unit for each member---between labor market work and household work. Members' allocation is a result of a non-cooperative game. Members of the household pool their total income and buy some consumption from it. There are also several household chores that, if not done, bring disutility to the household. Therefore, members need to allocate some fraction of their two units of time to the household chores. Neither member wishes to perform these chores and would rather work in the outside labor market. The goal of this section is to demonstrate how increases in wages for high-skilled labor can fundamentally alter the composition of skill acquisition choices in equilibrium. I will illustrate this using a simple parametric example. Consider a household $(m,w)$. Let the skill of member $i$ be $s_i$. The market wage for skill $s$ is $t_s$. Therefore, the wage of member $i$ is $t_i := t_{s_i}$. If member $i$'s allocation to the outside labor market is $e_i$, then the net household income is $y(s,e) := t\cdot e$, where $s := (s_m,s_w)$, $t:= (t_m,t_w)$, and $e := (e_m,e_w)$ are the skill, wage, and effort vectors. The quality of the household work as a result of their combined effort at home is $h(e):= 2 - e_m - e_w$. The utility of a member $i$ is given by, 
\begin{align*}
    U_i(s, e) = u(y(s,e), h(e)) - g(1-e_i).
\end{align*}
Essentially, the households treat both, their income and the level of household chores as public goods, while the cost, $g(\cdot)$, of their work at home is private. We assume that $u_1 > 0, u_{11} \le 0, u_2 > 0, u_{22} \le 0$. Moreover, the cost $g(\cdot)$ is assumed to be differentiable and convex. 

Members choose their allocations, $(e_i, e_j)$ in a non-cooperative manner taking the other person's choice as given. Let us write the agents' first order condition (assuming interior allocation choices): 
\begin{align}
    &u_1(y(s,e), h(e))t_i - u_2(y(s,e),h(e))+ g'(1-e_i) = 0\nonumber \\
    \implies & u_1(y(s,e),h(e))(t_i - t_j) = g'(1-e_j) - g'(1-e_i) \label{Equation: example, wage and microfoundation}
\end{align}
Therefore, if the players' choices are interior in equilibrium, then $t_i > t_j \implies e_i > e_j$ whenever $g(\cdot)$ is strictly convex.\footnote{In fact, this also implies that if $g(\cdot)$ were affine, then the allocation choices cannot be interior if the wages are unequal.} Thus, if the household consists of two members with unequal skills---and therefore unequal wages---then we will have specialization in equilibrium: the more skilled member will work more in the professional market than the one with lower skills. This observation is at the core of the following numerical example that illustrates how increasing wages for high-skilled workers can give rise to asymmetric equilibria. 

\bexam\label{Example: increasing wages and asymmetric} Consider a household consisting of $(m,w)$. The utility of member $i$ is $U(y(s,e),h(e)) = K \cdot [\a \log(t\cdot e) + (1-\a) \log(2 - e_m - e_w)] - \frac12 (1-e_i)^2$, where $K =8$ and $\a =0.6$. Let $t_l = 2$. Consider the wage for high-skills going from $t_h = 3 $ to $\hat t_h = 5$. The cost of acquiring skills for an agent with type $x$ is $\cost(x) = c + x$, where $c= 0.25$.
\eexam

When two members of skill $s$ and $s'$ match, they choose their effort choices non-cooperatively. Using \eqref{Equation: example, wage and microfoundation}, we solve this numerically for each pair of skills $(h,h), (h,l), (l,h), (l,l)$ to obtain $\phh,\phl,\plh,\pll$ corresponding to $t_h =3$ and $\phhhat,\phlhat,\plhhat,\pllhat$ corresponding to $t_h =5$. We provide a detailed working of this in Appendix~\ref{Section: example 2 derivation}, but the parameters are as follows:
\begin{align*}
    &\phh = 5.3547, \phl = 5.2114, \plh = 4.8495, \pll = 3.4085, \text{ and } \\
    &\phhhat = 7.8067, \phlhat = 7.7253, \plhhat = 7.2253, \pllhat = 3.4085.
\end{align*}

In the environment with $t_h = 3$, the unique equilibrium is symmetric with $\ttsym \approx 0.395$: the skill premium is too low to support any asymmetric equilibrium. When $t_h = 5$, the symmetric equilibrium is still SYM AM, now at the higher cutoff $\ttsym \approx 0.666$---as the premium rises, both genders invest more. But it is no longer alone: an asymmetric FIOS equilibrium with $\ttlb \approx 0.041$ becomes available alongside it, in which all men invest while only about $4\%$ of women do (the mirror image, with the genders swapped, is also an equilibrium). In the FIOS equilibrium, low-skilled women meet only high-skilled men on path, so the acceptance constraint \eqref{Equation: AM L accepts L} is vacuous for women, while it is verified to hold for the investing side. Thus, as wages rise, the economy moves from a regime in which about $40\%$ of each gender invest symmetrically to one in which that symmetric outcome coexists with a starkly asymmetric one---one gender investing fully, the other barely at all. Despite no asymmetry in the payoff or technological features of the underlying environment, the equilibrium set can exhibit stark asymmetries between the two genders as the skill premium rises.

\medskip

The pattern displayed in Example~\ref{Example: increasing wages and asymmetric}---a low-skill-premium regime in which the symmetric equilibrium is the only outcome, and a high-skill-premium regime in which an asymmetric (FIOS) equilibrium becomes available alongside the symmetric one---is not specific to the Cobb-Douglas example. The next definition isolates a broad class of household games for which the same comparative static holds.

\bdefn\label{Definition: PIH}
A \emph{pooled-income household (PIH) game} is the simultaneous-move game in which two members $i \in \{m, w\}$ with wages $(t_m, t_w)$ choose efforts $e_i \in [0, 1]$ to maximize
\begin{align*}
    U_i(e_m, e_w; t_m, t_w) = \psi(t_m e_m + t_w e_w,\, 2 - e_m - e_w) - g(1 - e_i),
\end{align*}
where
\begin{itemize}
    \item $\psi: \mathbb{R}_{++}^2 \to \mathbb{R}$ is twice continuously differentiable, strictly increasing in both arguments, weakly concave, and satisfies $\lim_{y \to \infty} \psi(y, h) = \infty$ for every fixed $h > 0$;
    \item $g: [0, 1] \to \mathbb{R}$ is twice continuously differentiable, strictly increasing, and strictly convex.
\end{itemize}
We assume $\psi$ and $g$ are such that the game has a \emph{unique} Nash equilibrium.\footnote{\citet{rosen1965existence} provides a sufficient condition for a class of games. For the Cobb-Douglas specification used in our example, uniqueness is straightforward to verify directly.} The match payoff $\phi(s, s')$ at skills $(s, s')$ is the value $U_i$ takes (for agent $i$ with skill $s$) at this equilibrium, with wages $(t_s, t_{s'})$.
\edefn

The Cobb-Douglas case in Example~\ref{Example: increasing wages and asymmetric} corresponds to $\psi(y, h) = K[\a \log y + (1-\a) \log h]$ and $g(z) = \tfrac{1}{2} z^2$.

In what follows, we give sufficient conditions on the asymptotic behavior of the induced match payoffs $\phi(s, s'; t_h)$ for a PIH game to produce the comparative-static phenomenon: a single wage threshold $t^*$ below which only the symmetric equilibrium exists and above which an asymmetric (FIOS) equilibrium becomes available alongside it. We refer to environments satisfying these conditions as displaying \emph{Increasing Submodularity in High Skills} (ISHS).

\bdefn[Increasing Submodularity in High Skills]\label{Definition: ISHS environment}
A marriage-market environment derived from a PIH game with $t_l > 0$ fixed, affine cost $\cost(x) = x + c$, and parameters $r, \lambda, c > 0$ \emph{displays Increasing Submodularity in High Skills} (ISHS) if the induced match payoffs $\phi(s, s'; t_h)$ satisfy:
\begin{enumerate}
    \item[(C1)] $\De(t_h) \to -\infty$ and $\De_h(t_h) \to \infty$ as $t_h \to \infty$;
    \item[(C2)] $\liminf_{t_h \to \infty} \ttsym(t_h) > 0$;
    \item[(C3)] Assumption~\ref{Assumption: payoff ranking} holds for all $t_h \ge t_l$, and there exists $T < \infty$ such that for all $t_h \ge T$, $\ttlb(t_h) \in (0, 1)$ and the FIOS profile $(1, \ttlb(t_h))$ satisfies \eqref{Equation: AM H accepts L} on path;
    \item[(C4)] $t_h \mapsto 1 + \lmr\De(t_h)$ is single-crossing on $[t_l, \infty)$: it changes sign at most once, and the change is from positive to negative.
\end{enumerate}
\edefn

In words, (C1) says that as the high-skill wage grows, the marriage payoffs become increasingly submodular: the asymmetric matches $(\phl, \plh)$ outpace the symmetric matches $(\phh, \pll)$ without bound. (C2) says that the symmetric equilibrium remains non-trivial: its investment cutoff stays bounded away from zero no matter how high the skill premium is. (C3) says that once the wage premium is large enough, the FIOS equilibrium structure is supported by the underlying payoffs---its cutoff $\ttlb(t_h)$ lies in $(0, 1)$, and the H-skilled side's acceptance constraint holds. (C4) is a single-crossing condition: once the payoffs become submodular enough to support FIOS ($1 + \lmr\De \le 0$), they stay so as the premium rises further. It is what makes ``FIOS becomes available above a threshold'' a genuine threshold statement rather than merely an eventual one; (C1) alone gives only the eventual sign of $1 + \lmr\De$, not the absence of a later return to positivity.

The Cobb-Douglas environment of Example~\ref{Example: increasing wages and asymmetric} displays ISHS. In particular (C4) is not vacuous there: in the corner regime ($t_h$ large) the $\log t_h$ terms of the $(H,H)$, $(H,L)$, and $(L,H)$ matches combine to give $\De(t_h) \sim -K\a\log t_h$, which is strictly decreasing, so $1 + \lmr\De$ is eventually strictly decreasing; in the calibration of Example~\ref{Example: increasing wages and asymmetric} it is monotone across its single zero at $t^* \approx 3.51$ (verified numerically).

\bprop\label{Proposition: parametric comparative static}
Let the marriage-market environment be derived from a PIH game and assume it displays ISHS (Definition~\ref{Definition: ISHS environment}). Then there is a wage threshold $t^* \in (t_l, \infty)$ such that a FIOS equilibrium coexists with the symmetric equilibrium for all $t_h > t^*$; while for every $t_h$ with $1 + \lmr\De(t_h) > 0$---in particular, for all sufficiently small $t_h$---no asymmetric equilibrium exists, so the symmetric equilibrium is the unique equilibrium.\footnote{Both asymmetric (AM) forms require $1 + \lmr\De < 0$ (the FIOS and NIOS investment ICs; see the proof of Proposition~\ref{Proposition: fragility selection}). In the ISHS calibration of Example~\ref{Example: increasing wages and asymmetric}, $1 + \lmr\De$ changes sign exactly at $t^* \approx 3.51$, so below $t^*$ the symmetric SYM AM equilibrium is the only equilibrium and above it a FIOS equilibrium coexists with it.} Suppose in addition that $\De_h'(t_h) \ge |\De'(t_h)|$,\footnote{This condition is satisfied for the Cobb-Douglas case in Example~\ref{Example: increasing wages and asymmetric}, where $\De_h'(t_h) = -\De'(t_h) = K\a/t_h$ once the $(H,L)$ household fully specializes ($t_h \gtrsim 4.4$ in this calibration), so $\De_h' = |\De'|$ there; for smaller $t_h$ the strict inequality $\De_h' > |\De'|$ holds (verified numerically). The hypothesis uses only the inequality.} and consider the range of wages over which the symmetric equilibrium is SYM AM. Then:
\begin{enumerate}
    \item the symmetric cutoff $\ttsym(t_h)$ is continuous and non-decreasing in $t_h$, so as the premium rises both genders invest more; and
    \item wherever the FIOS equilibrium exists, its cutoff lies strictly below the symmetric one, $\ttlb(t_h) < \ttsym(t_h)$, and is itself non-decreasing in $t_h$.
\end{enumerate}
\eprop

For the Cobb-Douglas example with $r = \lambda = 1$, $c = 0.25$, $K = 8$, $\a = 0.6$, $t_l = 2$, the threshold is $t^* \approx 3.51$: below it (as at $t_h = 3$) the symmetric SYM AM equilibrium is the only outcome, and above it (as at $t_h = 5$) an asymmetric FIOS equilibrium appears alongside it. The proof, in Appendix~\ref{Section: proof parametric comp static}, isolates the mechanism: rising $t_h$ drives $|\De|$ above the level at which the FIOS incentive constraints first become satisfiable. At still higher wages the symmetric equilibrium itself eventually turns selective---low-skilled agents begin rejecting one another, so SYM AM gives way to SYM SA (here at $t_{\Phi} \approx 6.11$)---but our interest is the emergence of the asymmetric equilibrium within the SYM AM range $t^* < t_h < t_\Phi$.

\paragraph{Empirical implications of Propositions~\ref{Proposition: equilibrium structure}--\ref{Proposition: parametric comparative static}.}
As the wage for high-skilled workers rises, the model's predictions divide into two phases. In the low-premium phase ($t_h < t^*$), where the payoffs are not yet submodular enough to admit an asymmetric equilibrium, the only equilibrium is symmetric and the investment cutoff rises continuously with the premium: both genders increase their investment in skill acquisition at the same rate. Empirically, this corresponds to broadly rising educational attainment, equal across genders.

In the high-premium phase, once $t_h$ crosses $t^*$ and while the symmetric equilibrium remains all-match, an asymmetric (FIOS) equilibrium becomes available alongside it. Under the fragility refinement of Section~\ref{Section: fragility}, this symmetric all-match equilibrium is fragile while the asymmetric one is not. The asymmetric equilibrium carries a sharp empirical signature: one gender invests in skill acquisition fully while the other invests substantially less and specializes in domestic production---a widening gender gap in education and labor-market participation that arises without any change in underlying preferences or technology, purely as the skill premium crosses the threshold. The model thus predicts that a rising skill premium can open up gender asymmetries exactly where none were present at lower premia.

\section{Conclusion}\label{Section: conclusion}

We have studied a canonical search-and-matching model with non-transferable utility in which, despite a fully symmetric environment, asymmetries in skill investment across genders can arise. Moreover, wherever a symmetric all-match equilibrium coexists with an asymmetric one, the former is fragile in the sense of \cite{onuchic2023signaling} while the latter is not, suggesting that the asymmetric outcomes are the more robust ones once they become available. Finally, the underlying changes in match payoffs can be microfounded through a household problem, allowing us to perform comparative statics on more primitive objects of the economy, such as labor-market wages.

The notion of fragility---which drives the selection argument in this paper---is, as \cite{onuchic2023signaling} themselves note, ``a mathematical construct, which may or may not have a bearing on actual equilibrium selection.'' I do not wish to claim that the asymmetric patterns of skill investment observed across genders in many settings are exclusively due to the fragility of their symmetric counterparts. What the notion does suggest, however, is that even minor asymmetries---or antisymmetries, more precisely---in investments arising from factors outside the model can be amplified at the symmetric equilibrium, rendering it unstable, while the FIOS and NIOS equilibria do not suffer from this defect.

Read alongside Propositions~\ref{Proposition: equilibrium structure}--\ref{Proposition: parametric comparative static}, the model illustrates three things: \emph{(i)} in an entirely symmetric environment, asymmetric investments in skill acquisition can arise in equilibrium; \emph{(ii)} such asymmetric equilibria are robust to small antisymmetric perturbations, whereas the symmetric all-match equilibria with which they coexist are fragile (Proposition~\ref{Proposition: fragility selection}); and \emph{(iii)} the shift from symmetric to asymmetric equilibria can be triggered by rising skill premia, which in turn can be linked to readily observable economic variables such as wages in the high-skill sector. In short, somewhat paradoxically, rising skill premia may end up amplifying or triggering gaps in skill investments across genders rather than mitigating them.

\bibliography{ref_matching.bib}

@article{legros2007beauty,
  author    = {Legros, Patrick and Newman, Andrew F.},
  title     = {Beauty Is a Beast, Frog Is a Prince: Assortative Matching 
               with Nontransferabilities},
  journal   = {Econometrica},
  year      = {2007},
  volume    = {75},
  number    = {4},
  pages     = {1073--1102},
  doi       = {10.1111/j.1468-0262.2007.00780.x}
}

@article{eeckhout1999,
  author    = {Eeckhout, Jan},
  title     = {Bilateral Search and Vertical Heterogeneity},
  journal   = {Econometrica},
  year      = {1999},
  volume    = {67},
  number    = {6},
  pages     = {1291--1311},
  doi       = {10.1111/1468-0262.00082}
}

@article{iyigun2007,
  author    = {Iyigun, Murat and Walsh, Randall P.},
  title     = {Building the Family Nest: Premarital Investments, Marriage 
               Markets, and Spousal Allocations},
  journal   = {Review of Economic Studies},
  year      = {2007},
  volume    = {74},
  number    = {2},
  pages     = {507--535},
  doi       = {10.1111/j.1467-937X.2007.00425.x}
}

@article{lafortune2013,
  author    = {Lafortune, Jeanne},
  title     = {Making Yourself Attractive: Pre-Marital Investments and the 
               Returns to Education in the Marriage Market},
  journal   = {American Economic Journal: Applied Economics},
  year      = {2013},
  volume    = {5},
  number    = {2},
  pages     = {151--178},
  doi       = {10.1257/app.5.2.151}
}

@article{greenwood2014,
  author    = {Greenwood, Jeremy and Guner, Nezih and Kocharkov, Georgi 
               and Santos, Cezar},
  title     = {Marry Your Like: Assortative Mating and Income Inequality},
  journal   = {American Economic Review},
  year      = {2014},
  volume    = {104},
  number    = {5},
  pages     = {348--353},
  doi       = {10.1257/aer.104.5.348}
}

@article{goldin2006,
  author    = {Goldin, Claudia},
  title     = {The Quiet Revolution That Transformed Women's Employment, 
               Education, and Family},
  journal   = {American Economic Review},
  year      = {2006},
  volume    = {96},
  number    = {2},
  pages     = {1--21},
  doi       = {10.1257/000282806777212350}
}

@article{atakan2026efficient,
  title={Efficient Investment, Search, and Sorting in Matching Markets},
  author={Atakan, Alp and Richter, Michael and Tsur, Matan},
  journal={Theoretical Economics},
  note={Forthcoming},
  year={2026}
}

@article{smith2006marriage,
  title={The marriage model with search frictions},
  author={Smith, Lones},
  journal={Journal of political Economy},
  volume={114},
  number={6},
  pages={1124--1144},
  year={2006},
  publisher={The University of Chicago Press}
}

@article{Chiappori1988,
  author    = {Pierre-Andr{\'e} Chiappori},
  title     = {Rational Household Labor Supply},
  journal   = {Econometrica},
  volume    = {56},
  number    = {1},
  pages     = {63--90},
  year      = {1988},
  publisher = {Econometric Society},
  doi       = {10.2307/1911842},
  url       = {https://doi.org/10.2307/1911842}
}

@article{bhaskar2023multidimensional,
  title={Multidimensional Premarital Investments with Imperfect Commitment},
  author={Bhaskar, Venkataraman and Li, Wenchao and Yi, Junjian},
  journal={Journal of Political Economy},
  volume={131},
  number={10},
  pages={2893--2919},
  year={2023},
  publisher={The University of Chicago Press Chicago, IL}
}

@article{chiappori2017partner,
  title={Partner choice, investment in children, and the marital college premium},
  author={Chiappori, Pierre-Andr{\'e} and Salani{\'e}, Bernard and Weiss, Yoram},
  journal={American Economic Review},
  volume={107},
  number={8},
  pages={2109--2167},
  year={2017},
  publisher={American Economic Association 2014 Broadway, Suite 305, Nashville, TN 37203}
}

@article{chiappori2009investment,
  title={Investment in schooling and the marriage market},
  author={Chiappori, Pierre-Andr{\'e} and Iyigun, Murat and Weiss, Yoram},
  journal={American Economic Review},
  volume={99},
  number={5},
  pages={1689--1713},
  year={2009},
  publisher={American Economic Association}
}

@article{shimer2001matching,
  title={Matching, search, and heterogeneity},
  author={Shimer, Robert and Smith, Lones},
  journal={Topics in Macroeconomics},
  volume={1},
  number={1},
  pages={153460131010},
  year={2001},
  publisher={De Gruyter}
}

@article{shimer2000assortative,
  title={Assortative matching and search},
  author={Shimer, Robert and Smith, Lones},
  journal={Econometrica},
  volume={68},
  number={2},
  pages={343--369},
  year={2000},
  publisher={Wiley Online Library}
}

@article{chade2022risky,
  title={Risky matching},
  author={Chade, Hector and Lindenlaub, Ilse},
  journal={The Review of Economic Studies},
  volume={89},
  number={2},
  pages={626--665},
  year={2022},
  publisher={Oxford University Press}
}

@article{adachi2003search,
  title={A search model of two-sided matching under nontransferable utility},
  author={Adachi, Hiroyuki},
  journal={Journal of Economic Theory},
  volume={113},
  number={2},
  pages={182--198},
  year={2003},
  publisher={Elsevier}
}

@article{shapley1971assignment,
  title={The assignment game I: The core},
  author={Shapley, Lloyd S and Shubik, Martin},
  journal={International Journal of game theory},
  volume={1},
  number={1},
  pages={111--130},
  year={1971},
  publisher={Springer}
}

@article{mailath2013pricing,
  title={Pricing and investments in matching markets},
  author={Mailath, George J and Postlewaite, Andrew and Samuelson, Larry},
  journal={Theoretical Economics},
  volume={8},
  number={2},
  pages={535--590},
  year={2013},
  publisher={Wiley Online Library}
}

@article{peters2007pre,
  title={The pre-marital investment game},
  author={Peters, Michael},
  journal={Journal of Economic Theory},
  volume={137},
  number={1},
  pages={186--213},
  year={2007},
  publisher={Elsevier}
}

@article{peters2002competing,
  title={Competing Premarital Investments},
  author={Peters, Michael and Siow, Aloysius},
  journal={Journal of Political Economy},
  volume={110},
  number={3},
  pages={592--608},
  year={2002}
}

@article{galichon2022cupid,
  title={Cupid's Invisible Hand: Social Surplus and Identification in Matching Models},
  author={Galichon, Alfred and Salani{\'e}, Bernard},
  journal={Review of Economic Studies},
  volume={89},
  number={5},
  pages={2600--2629},
  year={2022}
}

@article{choo2006who,
  title={Who Marries Whom and Why},
  author={Choo, Eugene and Siow, Aloysius},
  journal={Journal of Political Economy},
  volume={114},
  number={1},
  pages={175--201},
  year={2006}
}

@article{cole2001efficient,
  title={Efficient non-contractible investments in large economies},
  author={Cole, Harold L and Mailath, George J and Postlewaite, Andrew},
  journal={Journal of Economic Theory},
  volume={101},
  number={2},
  pages={333--373},
  year={2001},
  publisher={Elsevier}
}

@article{noldeke2015investment,
  title={Investment and competitive matching},
  author={N{\"o}ldeke, Georg and Samuelson, Larry},
  journal={Econometrica},
  volume={83},
  number={3},
  pages={835--896},
  year={2015},
  publisher={Wiley Online Library}
}

@article{chiappori2020theory,
  title={The theory and empirics of the marriage market},
  author={Chiappori, Pierre-Andr{\'e}},
  journal={Annual Review of Economics},
  volume={12},
  number={1},
  pages={547--578},
  year={2020},
  publisher={Annual Reviews}
}

@article{bhaskar2016marriage,
  title={Marriage as a rat race: Noisy premarital investments with assortative matching},
  author={Bhaskar, Venkataraman and Hopkins, Ed},
  journal={Journal of Political Economy},
  volume={124},
  number={4},
  pages={992--1045},
  year={2016},
  publisher={University of Chicago Press Chicago, IL}
}

@article{gale1962college,
  title={College admissions and the stability of marriage},
  author={Gale, David and Shapley, Lloyd S},
  journal={The American Mathematical Monthly},
  volume={69},
  number={1},
  pages={9--15},
  year={1962},
  publisher={Taylor \& Francis}
}

@article{bertrand2015gender,
  title={Gender identity and relative income within households},
  author={Bertrand, Marianne and Kamenica, Emir and Pan, Jessica},
  journal={The Quarterly Journal of Economics},
  volume={130},
  number={2},
  pages={571--614},
  year={2015},
  publisher={MIT Press}
}

@article{zhang2021investment,
  title={An investment-and-marriage model with differential fecundity: On the college gender gap},
  author={Zhang, Hanzhe},
  journal={Journal of Political Economy},
  volume={129},
  number={5},
  pages={1464--1486},
  year={2021},
  publisher={The University of Chicago Press Chicago, IL}
}

@article{siow1998differntial,
  title={Differntial fecundity, markets, and gender roles},
  author={Siow, Aloysius},
  journal={Journal of Political Economy},
  volume={106},
  number={2},
  pages={334--354},
  year={1998},
  publisher={The University of Chicago Press}
}

@article{low2024human,
  title={The human capital--reproductive capital trade-off in marriage market matching},
  author={Low, Corinne},
  journal={Journal of Political Economy},
  volume={132},
  number={2},
  pages={552--576},
  year={2024},
  publisher={The University of Chicago Press Chicago, IL}
}

@article{becker1973theory,
  title={A theory of marriage: Part I},
  author={Becker, Gary S},
  journal={Journal of Political economy},
  volume={81},
  number={4},
  pages={813--846},
  year={1973},
  publisher={The University of Chicago Press}
}

@article{rosen1965existence,
  title={Existence and uniqueness of equilibrium points for concave $n$-person games},
  author={Rosen, J. B.},
  journal={Econometrica},
  volume={33},
  number={3},
  pages={520--534},
  year={1965}
}

@book{goldin2008race,
  author    = {Goldin, Claudia and Katz, Lawrence F.},
  title     = {The Race Between Education and Technology},
  publisher = {Belknap Press of Harvard University Press},
  year      = {2008},
  address   = {Cambridge, MA}
}

@incollection{acemoglu2011skills,
  author    = {Acemoglu, Daron and Autor, David},
  title     = {Skills, Tasks and Technologies: Implications for Employment and Earnings},
  booktitle = {Handbook of Labor Economics},
  editor    = {Ashenfelter, Orley and Card, David},
  publisher = {Elsevier},
  year      = {2011},
  volume    = {4B},
  pages     = {1043--1171}
}

@article{burdett1997marriage,
  author    = {Burdett, Kenneth and Coles, Melvyn G.},
  title     = {Marriage and Class},
  journal   = {Quarterly Journal of Economics},
  year      = {1997},
  volume    = {112},
  number    = {1},
  pages     = {141--168}
}

@article{hadfield1999coordination,
  author  = {Hadfield, Gillian K.},
  title   = {A Coordination Model of the Sexual Division of Labor},
  journal = {Journal of Economic Behavior and Organization},
  year    = {1999},
  volume  = {40},
  number  = {2},
  pages   = {125--153}
}

@article{francois1998gender,
  author  = {Francois, Patrick},
  title   = {Gender Discrimination without Gender Difference: Theory and Policy Responses},
  journal = {Journal of Public Economics},
  year    = {1998},
  volume  = {68},
  number  = {1},
  pages   = {1--32}
}

@article{onuchic2023signaling,
  title={Signaling and Discrimination in Collaborative Projects},
  author={Onuchic, Paula and Ray, Debraj},
  journal={American Economic Review},
  volume={113},
  number={1},
  pages={210--252},
  year={2023}
}

@article{lundberg1993separate,
  title={Separate Spheres Bargaining and the Marriage Market},
  author={Lundberg, Shelly and Pollak, Robert A.},
  journal={Journal of Political Economy},
  volume={101},
  number={6},
  pages={988--1010},
  year={1993}
}

\appendix
\newpage
\section{Appendix: Proofs}

\subsection{Symmetry of AM equilibria under supermodular payoffs}

The following auxiliary result is referenced in Section~\ref{Section: AM equilibria analysis} to justify the focus on submodular payoffs. It says that under supermodular payoffs, no AM equilibrium can exhibit asymmetric investment cutoffs across genders; in particular, the FIOS and NIOS asymmetric structures are ruled out in this case. Combined with the analysis under submodular payoffs in the main text, this result confirms that asymmetric outcomes require submodularity of the match payoff function.

\bprop\label{Proposition: unique symmetric AM eqm if SM}
If $\payoff(\cdot, \cdot)$ is supermodular ($\De \ge 0$), then all AM equilibria are symmetric, i.e., $\ttypem = \ttypew$.
\eprop

\bprf[Proof of Proposition~\ref{Proposition: unique symmetric AM eqm if SM}]
Let us split the analysis into three cases: 

\noindent\textbf{Case 1: }$ \ttype \in (0,1)^2$. 
Suppose that $\ttypem \neq \ttypew$. From \eqref{Equation: AM invest IC}, we have, 
\begin{align*}
    \valfn^\g_A(H; \ttype) - \valfn^\g_A(L; \ttype) &= \frac{\lm}{r + \lm}\bigg[ \ttypec_{\gt}(\underbrace{\phh + \pll - \phl - \plh}_{ \ge 0}) \\
    &\quad + \underbrace{\phl - \pll}_{ \ge 0 }\bigg] \\
    &= \cost(\ttypecg).
\end{align*}

Therefore, 
\begin{align*}
&\big[\valfn^m_A(H; \ttype) - \valfn^m_A(L; \ttype) \big] - \big[\valfn^w_A(H; \ttype) - \valfn^w_A(L; \ttype) \big] \\
&= \frac{\lm}{r + \lm} (\ttypew - \ttypem) \De \\
&= \cost(\ttypem) - \cost(\ttypew)
\end{align*}
Therefore, if $\ttypem > \ttypew$, then $LHS \le 0 < RHS$ (with $LHS = 0$ only at the modular edge $\De = 0$, where the equality $LHS = RHS$ then forces $\ttypem = \ttypew$), a contradiction.

\noindent\textbf{Case 2: }$ \ttypem \in [0,1)$ and $\ttypew = 1$. 
Then, 
\begin{align*}
    \valfn^m_A(H;\ttype) - \valfn^m_A(L;\ttype) = & \lmr[ \ttypew \De + \De_h]\le \cost(\ttypem) \\
    \valfn^w_A(H;\ttype) - \valfn^w_A(L;\ttype) = & \lmr[ \ttypem \De + \De_h] \ge  \cost(\ttypew) 
\end{align*}

Therefore, 
\begin{align*}
    &[\valfn^m_A(H;\ttype) - \valfn^m_A(L;\ttype)] - [\valfn^w_A(H;\ttype) - \valfn^w_A(L;\ttype) ] \\
    &=  \lmr (\ttypew-\ttypem) \De = \lmr (1-\ttypem) \De \le  \cost(\ttypem) - \cost(1) 
\end{align*}
However, $LHS \ge 0 > RHS$, a contradiction. 

\noindent\textbf{Case 3: }$ \ttypem \in (0,1]$ and $\ttypew = 0$. 
Then, 
\begin{align*}
    \valfn^m_A(H;\ttype) - \valfn^m_A(L;\ttype) = & \lmr[ \ttypew \De + \De_h]\ge \cost(\ttypem) \\
    \valfn^w_A(H;\ttype) - \valfn^w_A(L;\ttype) = & \lmr[ \ttypem \De + \De_h] \le  \cost(\ttypew) 
\end{align*}

Therefore, 
\begin{align*}
    &[\valfn^m_A(H;\ttype) - \valfn^m_A(L;\ttype)] - [\valfn^w_A(H;\ttype) - \valfn^w_A(L;\ttype) ] \\
    &=  \lmr (\ttypew-\ttypem) \De = -\lmr \ttypem \De \ge  \cost(\ttypem) - \cost(0) 
\end{align*}
However, $LHS \le 0 < RHS$, a contradiction. 
\eprf 

\subsection{Proof of Proposition~\ref{Proposition: equilibrium structure}}

\bprf[Proof of Proposition~\ref{Proposition: equilibrium structure}]
For the reader's convenience, recall the payoff aggregates
\begin{align*}
    \De &:= \phh + \pll - \phl - \plh, \\
    \De_h &:= \phl - \pll, \\
    \eta &:= \plh - \pll,
\end{align*}
and the closed-form cutoff candidates:
\begin{align*}
    \ttsym = \frac{\lmr \De_h - c}{1 - \lmr \De}, \qquad
    \ttlb = \lmr(\De + \De_h) - c, \qquad
    \ttub = \lmr \De_h - c.
\end{align*}
Throughout, $\De \le 0$, Assumption~\ref{Assumption: payoff ranking} (which gives $\De_h \ge \eta \ge 0$ and $\phl \ge \pll$), and the maintained condition \eqref{Equation: no PAM} hold.

\medskip
\noindent \textbf{Proof of Part 1 (symmetric equilibrium).}
Throughout this proof, we treat the generic interior case in which the unclipped formula for $\ttsym$ in \eqref{Equation: cutoff candidates} lies in $(0, 1)$; the corner cases $\ttsym \in \{0, 1\}$ correspond to the all-L or all-H profiles whose acceptance constraints are vacuous and are easily verified.

At any symmetric AM profile $(\ttype, \ttype)$, \eqref{Equation: AM invest IC} reduces to
\begin{align*}
    \lmr[\ttype \De + \De_h] = \ttype + c,
\end{align*}
whose unique interior solution is $\ttype = \ttsym$ (the LHS is non-increasing and the RHS strictly increasing in $\ttype$, since $\De \le 0$). The H-L acceptance constraint \eqref{Equation: AM H accepts L} at $(\ttsym, \ttsym)$ follows from the L-L constraint \eqref{Equation: AM L accepts L} via submodularity ($\phh - \phl \le \eta$) and $\phl \ge \pll$:
\begin{align*}
    \lm \ttsym (\phh - \phl) \le \lm \ttsym \eta \le r \pll \le r \phl.
\end{align*}
Thus SYM AM's existence is determined by the L-L acceptance condition alone.

\smallskip
\noindent \emph{Case A: $\lm \ttsym \eta \le r \pll$.} The profile $(\ttsym, \ttsym)$ constitutes the unique SYM AM equilibrium. The SA L-rejection condition \eqref{Equation: SA L-L reject}, $\lm \ttype \eta > r \pll$, is the strict negation of the L-L acceptance, so no SYM SA can coexist \emph{at the cutoff $\ttsym$}; that no SYM SA exists at any other cutoff either is established at the end of this proof.

\smallskip
\noindent \emph{Case B: $\lm \ttsym \eta > r \pll$.} SYM AM does not exist. Define the threshold cutoff
\begin{align*}
    \ttstar := \frac{r \pll}{\lm \eta} \in (0, \ttsym),
\end{align*}
at which the AM L-L acceptance just binds. At $\ttstar$, the L-type AM and SA values coincide and both equal $\pll$:
\begin{align*}
    \valfn^\g_A(L; \ttstar) &= \frac{\lm[\ttstar \plh + (1-\ttstar)\pll]}{r + \lm}
        = \frac{\lm \ttstar \eta + \lm \pll}{r+\lm}  = \pll, \\
    \valfn^\g_S(L; \ttstar) &= \frac{\lm \ttstar \plh}{r + \lm \ttstar} = \pll,
\end{align*}
where for the second line we use $r + \lm \ttstar = \lm \ttstar \plh / \pll$ (rearrangement of $r\pll = \lm \ttstar \eta$). Since H always matches at rate $\lm$ under both AM and SA, $\valfn^\g_S(H; \ttype) = \valfn^\g_A(H; \ttype)$ for all $\ttype$. For each regime, define the net gain to investing for the cutoff type---that is, the value of being H minus the value of being L minus the investment cost---as
\begin{align*}
    F_S(\ttype) &:= \valfn^\g_S(H; \ttype) - \valfn^\g_S(L; \ttype) - (\ttype + c), \\
    F_A(\ttype) &:= \valfn^\g_A(H; \ttype) - \valfn^\g_A(L; \ttype) - (\ttype + c) = \lmr[\ttype \De + \De_h] - (\ttype + c).
\end{align*}
Each $F_\sigma(\ttype) = 0$ is the symmetric investment indifference condition in regime $\sigma \in \{S, A\}$ at cutoff $\ttype$. Since $\valfn^\g_S(L; \ttstar) = \valfn^\g_A(L; \ttstar) = \pll$ and $\valfn^\g_S(H; \ttype) = \valfn^\g_A(H; \ttype)$, we have $F_S(\ttstar) = F_A(\ttstar)$. Since $F_A$ is strictly decreasing in $\ttype$ (as $\De \le 0$) and $F_A(\ttsym) = 0$ with $\ttstar < \ttsym$,
\begin{align*}
    F_S(\ttstar) = F_A(\ttstar) > F_A(\ttsym) = 0.
\end{align*}
At the upper endpoint, $F_S(1) = \lm(\phh - \plh)/(r+\lm) - 1 - c$. In the corner subcase $F_S(1) \ge 0$, the profile $(1,1)$ is itself a symmetric equilibrium (everyone invests). Otherwise, by the intermediate value theorem, $F_S$ has a zero $\ttsa \in (\ttstar, 1)$ satisfying \eqref{Equation: SA invest IC}. At this root, $\lm \ttsa \eta > \lm \ttstar \eta = r \pll$, so \eqref{Equation: SA L-L reject} holds strictly.

For uniqueness of $\ttsa$, note that
\begin{align*}
    F_S'(\ttype) = \frac{\lm(\phh - \phl)}{r + \lm} - \frac{\lm r \plh}{(r + \lm \ttype)^2} - 1
\end{align*}
is strictly increasing in $\ttype$ (the middle term is decreasing), so $F_S$ is convex. A convex function with $F_S(\ttstar) > 0$ and $F_S(1) < 0$ has exactly one zero on $(\ttstar, 1)$. Hence $\ttsa$ is unique.

Moreover, $\ttsa < \ttsym$. For $\ttype \in (\ttstar, 1)$ the L-type strictly prefers to reject low partners, so $\valfn^\g_S(L; \ttype) > \valfn^\g_A(L; \ttype)$ and hence $F_S(\ttype) < F_A(\ttype)$; in particular $F_S(\ttsym) < F_A(\ttsym) = 0$. Since $F_S(\ttstar) > 0$ and $\ttsa$ is the unique zero of $F_S$ on $(\ttstar, 1)$, the downward crossing satisfies $\ttsa \in (\ttstar, \ttsym)$.

It remains to verify that high-skilled agents accept low-skilled partners at $\ttsa$, as an SA configuration requires. A high type's acceptance constraint is $\lm \ttsa (\phh - \phl) \le r \phl$; since $\ttsa \le 1$, the maintained condition \eqref{Equation: no PAM} gives $\lm \ttsa (\phh - \phl) \le \lm (\phh - \phl) < r \phl$, so the constraint holds. Thus $(\ttsa, \ttsa)$ satisfies all the SA constraints and is a SYM SA equilibrium.

Finally, we argue that SYM AM and SYM SA cannot coexist, even at different cutoffs. Suppose SYM AM exists at $\ttsym$. Its L-L acceptance condition gives $\lm \ttsym \eta \le r \pll$, equivalently $\ttsym \le \ttstar$. Since $F_A$ is strictly decreasing and vanishes at $\ttsym$, we have $F_A(\ttstar) \le F_A(\ttsym) = 0$. Since L-type AM and SA values coincide at $\ttstar$ (shown above), $F_S(\ttstar) = F_A(\ttstar) \le 0$. By the convexity of $F_S$ established above, together with the generic case $F_S(1) < 0$, we have $F_S \le 0$ on $[\ttstar, 1]$. Hence $F_S$ admits no zero on $[\ttstar, 1]$ at which L-rejection ($\lm \ttype \eta > r \pll$, i.e., $\ttype > \ttstar$) holds, so no SYM SA equilibrium exists. The argument in Case B above showed the symmetric converse: when SYM SA exists, SYM AM does not. The symmetric equilibrium is therefore unique in either case.

\medskip
\noindent \textbf{Proof of Part 2 (form of asymmetric equilibria).}
\emph{At most one asymmetric AM equilibrium.} Suppose an interior asymmetric equilibrium $(\ttypem, \ttypew) \in (0,1)^2$ with $\ttypem \neq \ttypew$ exists. The investment ICs \eqref{Equation: AM invest IC} yield
\begin{equation}
\begin{split}
    \lmr [\ttypew \De + \De_h] &= \ttypem + c, \\
    \lmr [\ttypem \De + \De_h] &= \ttypew + c.
\end{split}
\label{Equation: interior asymmetric}
\end{equation}
Adding the two equations gives $\ttypem + \ttypew = 2 \ttsym$, while subtracting gives $(\ttypem - \ttypew)(1 + \lmr \De) = 0$. Since $\ttypem \neq \ttypew$, this requires the knife-edge condition $1 + \lmr \De = 0$, which we exclude. Hence no interior asymmetric AM equilibrium exists generically.

It follows that any asymmetric AM equilibrium has at least one cutoff at $\{0, 1\}$. Suppose without loss of generality that $\ttypem = 1$. The marginal woman's best response at men's cutoff $1$ is the unique solution to
\begin{align*}
    \ttypew(1) = \begin{cases}
        0 & \text{if } \lmr [\De + \De_h] \le c, \\
        1 & \text{if } \lmr [\De + \De_h] \ge 1 + c, \\
        \lmr [\De + \De_h] - c = \ttlb & \text{otherwise,}
    \end{cases}
\end{align*}
uniqueness following from strict monotonicity of $\lmr[\De + \De_h] - (y + c)$ in $y$. The candidate is therefore the FIOS profile $(1, \ttlb)$. For this candidate to be an equilibrium, the fully-investing gender's corner must be incentive-compatible against the deviation-optimal value of a low type (the fully-investing side carries no low types on path, so matching is all-match on path and the low type's acceptance is set by optimality): this is $\valfn_A(H;\ttlb)-(1+c)\ge\max\{\valfn_A(L;\ttlb),\valfn_S(L;\ttlb)\}$, i.e.\ $\min\{\lmr(\ttlb\De+\De_h),\,G(\ttlb)\}\ge 1+c$, which coincides with the all-match condition $\lmr(\ttlb\De+\De_h)\ge1+c$ exactly when $\lm\ttlb\,\eta\le r\pll$. Symmetrically, if $\ttypem = 0$, the unique candidate is the NIOS profile $(0, \ttub)$; here the all-low side is populated on path, and its L-L acceptance $\lm\ttub\,\eta\le r\pll$ is imposed directly. Up to gender permutation, FIOS and NIOS are the only candidates.

To rule out coexistence of FIOS and NIOS, suppose both $(1, \ttlb)$ and $(0, \ttub)$ are equilibria with $\ttlb, \ttub \in (0,1)$. Note $\ttub = \ttlb - \lmr \De > \ttlb$ since $\De < 0$. The women's ICs deliver the contradiction directly: FIOS women's indifference gives $\lmr[\De + \De_h] - c = \ttlb \ge 0$. Since $\ttub > \ttlb$ and $\De < 0$,
\begin{align*}
    \lmr [\De \ttub + \De_h] - c > \lmr [\De + \De_h] - c = \ttlb \ge 0,
\end{align*}
where the strict inequality uses $\De < 0$ and $\ttub < 1$, so $\De \ttub > \De$. But this contradicts NIOS men's IC $\lmr[\ttub \De + \De_h] \le c$.\footnote{At the boundaries the two profiles degenerate to mirror images of one another: $\ttlb = 0$ makes FIOS the profile $(1,0)$---one side fully invests, the other not at all---and $\ttub = 1$ makes NIOS the profile $(0,1)$, the same full-specialization configuration with the genders swapped. They are then not distinct coexisting equilibria, and the interior argument above is what rules out genuine coexistence.} Hence at most one asymmetric (AM) equilibrium exists.

\smallskip
\noindent \emph{When the symmetric equilibrium is SA, the asymmetric form must be FIOS.} We show that under the hypothesis $\lm \ttsym \eta > r \pll$ (the regime in which the symmetric equilibrium is SA), no NIOS equilibrium can exist. Combined with the argument above ruling out interior asymmetric profiles, this leaves FIOS as the only possible asymmetric form.

The NIOS profile $(0, \ttub)$ requires the L-L acceptance for L-skilled men (the all-low side) facing women's pool $\ttub$:
\begin{align*}
    \lm \ttub \eta \le r \pll.
\end{align*}
But $\ttub = (1 - \lmr \De) \ttsym \ge \ttsym$, since $1 - \lmr \De \ge 1$ (as $\De \le 0$). Therefore,
\begin{align*}
    \lm \ttub \eta \ge \lm \ttsym \eta > r \pll,
\end{align*}
contradicting the NIOS L-L constraint. Hence NIOS does not exist under failure of SYM AM. The boundary case $\ttub \ge 1$ (NIOS profile $(0,1)$) yields $\lm \eta \le r \pll$ for the L-L constraint at women's all-H pool; but $\lm \eta \ge \lm \ttsym \eta > r \pll$ (since $\ttsym \le 1$), again a contradiction.

\eprf

\subsection{Proof of Proposition~\ref{Proposition: higher skill premia open asymmetric}}
\bprf[Proof of Proposition~\ref{Proposition: higher skill premia open asymmetric}]
We first show that $\bpayoff$ admits no asymmetric AM equilibrium. By Proposition~\ref{Proposition: equilibrium structure} Part 2, any such equilibrium must take FIOS or NIOS form, so it suffices to rule out each.

\emph{No NIOS.} The NIOS profile $(0, \ttub)$ requires men's IC at type $0$, $\lmr[\ttub \De + \De_h] \le c$. Substituting $\ttub = \lmr \De_h - c$ and rearranging gives
\begin{align*}
    \left(\lmr \De_h - c\right) \left(1 + \lmr \De\right) \le 0.
\end{align*}
By (ii), $\lmr \De_h - c > 0$; by (i), $1 + \lmr \De > 0$. The product is therefore strictly positive, contradicting the inequality. Hence NIOS is not an equilibrium under $\bpayoff$.

\medskip

\emph{No FIOS.} The FIOS investing-side IC requires $\min\{\lmr(\ttlb\De+\De_h),G(\ttlb)\}\ge 1+c$; it suffices to violate the all-match term $\lmr[\ttlb \De + \De_h] \ge 1 + c$ (a fortiori the deviation-optimal IC, being a minimum, then also fails). Substituting $\ttlb = \lmr(\De + \De_h) - c$ and rearranging gives
\begin{align*}
    \left(\lmr \De_h - c\right)\left(1 + \lmr \De\right) \ge \left(1 - \lmr \De\right)\left(1 + \lmr \De\right).
\end{align*}
Under (i), $1 + \lmr \De > 0$, so this reduces to $\lmr \De_h - c \ge 1 - \lmr \De$, equivalently $\ttlb = \lmr(\De + \De_h) - c \ge 1$. But submodularity ($\De \le 0$) gives $\ttlb \le \ttub = \lmr \De_h - c < 1$ by (ii), a contradiction. Hence FIOS is not an equilibrium under $\bpayoff$ either.

Combined with the proposition's premise, the unique equilibrium of $\bpayoff$ is the symmetric AM equilibrium.

\medskip

Now, let us turn to constructing $\bpayoffh$ that exhibits a higher skill premium relative to $\bpayoff$ and admits a FIOS equilibrium (alongside the symmetric equilibrium guaranteed by Proposition~\ref{Proposition: equilibrium structure}). In particular, we will choose $\phhhat, \phlhat,\plhhat$ larger than $\phh,\phl,\plh$ in a way that supports a FIOS equilibrium. Let us rewrite the constraints of a FIOS equilibrium.
\begin{align*}
    \lmr (\De + \De_h) -c \in& (0,1). \\
    \lmr (\ttlb \De + \De_h) \ge&  1 + c\\
    \lmr ( \ttlb \phhhat + (1-\ttlb) \phlhat) \le& \phlhat
\end{align*}
Of the three constraints above, our construction will keep $\phhhat \approx \phlhat$, so the third (H-L acceptance) constraint is automatically satisfied and we focus on the first two.

We keep $\pllhat = \pll$ unchanged and choose $\plhhat = a$ sufficiently larger than $\pll$, with $\phhhat \approx \phlhat = b > a$. Then $\widehat{\De} \approx \pll - a$ and $\widehat{\De_h} \approx b - \pll$. As before, \eqref{Equation: AM invest IC} for a FIOS equilibrium under $\bpayoffh$ rearranges to
\begin{align}
    \left(\lmr \widehat{\De_h} - c\right) \left(1 + \lmr \widehat{\De}\right) \ge \left(1 - \lmr \widehat{\De}\right) \left(1 + \lmr \widehat{\De}\right)\label{Equation: AM invest IC-construction}
\end{align}
This requires $1 + \lmr \widehat{\De} \le 0$, equivalently $\lmr (a - \pll) \ge 1$. As $a$ increases, $|\widehat{\De}|$ increases, so the inequality is satisfied for $a$ sufficiently larger than $\pll$. We also need $\widehat{\ttlb} = \lmr(\widehat{\De} + \widehat{\De_h}) - c \approx \lmr(b-a) - c$ to lie in $(0,1)$. Given any admissible $a$, $\widehat{\ttlb} < 1$ puts an upper bound on $b$ and $\widehat{\ttlb} > 0$ puts a lower bound. Once these hold, \eqref{Equation: AM invest IC-construction} reduces, under $1 + \lmr \widehat{\De} \le 0$, to $\widehat{\ttlb} \le 1$, which is exactly the upper bound on $b$. Finally, since the fully-investing side carries no low types on path, its corner IC is the deviation-optimal condition $\min\{\lmr(\widehat\ttlb\widehat\De+\widehat\De_h),G(\widehat\ttlb)\}\ge 1+c$ (see the definition of the FIOS profile). The all-match term holds by the reduction just given; the SA term $G(\widehat\ttlb)=\lmr[\widehat\ttlb\,\phhhat+(1-\widehat\ttlb)\phlhat]-\lm\widehat\ttlb\,\plhhat/(r+\lm\widehat\ttlb)\approx \lmr b - \lm\widehat\ttlb\, a/(r+\lm\widehat\ttlb)$ exceeds $1+c$ once $b$ is taken large enough (with the gap $b-a$, hence $\widehat\ttlb$, held fixed), which is compatible with the bounds above. Therefore, choosing $a$ large enough, then $b$ large enough with $b-a$ set so that $\widehat{\ttlb} \in (0,1)$, guarantees that a FIOS equilibrium exists under $\bpayoffh$. This completes the construction.
\eprf

\subsection{Proof of Proposition~\ref{Proposition: rising investment}}
\bprf[Proof of Proposition~\ref{Proposition: rising investment}]
Write $\ttsym = (\lmr\De_h - c)/(1 - \lmr\De) =: N/D$, with $D = 1 - \lmr\De > 0$ and $N = \lmr\De_h - c = \ttsym\, D > 0$. Differentiating, $\ttsym' = \lmr(D\,\De_h' + N\,\De')/D^2$, so $\ttsym' \ge 0$ if and only if $\De_h' \ge -\ttsym\,\De'$. Since $\ttsym \in (0,1)$ and $\De_h' \ge 0$: if $\De' \ge 0$ the right-hand side is non-positive; if $\De' < 0$ then $-\ttsym\,\De' = \ttsym\,|\De'| \le |\De'| \le \De_h'$. In either case $\ttsym' \ge 0$.
\eprf

\subsection{Proof of Proposition~\ref{Proposition: fragility selection}}

\bprf[Proof of Proposition~\ref{Proposition: fragility selection}]
Coexistence of FIOS at $(1,\ttlb)$ with interior $\ttlb \in (0,1)$ requires the men's investment IC at cutoff 1. The deviation-optimal form of this IC is $\min\{\lmr(\De\ttlb+\De_h),G(\ttlb)\}\ge1+c$ (see the definition of the FIOS profile); in particular it implies the all-match term $\lmr(\De\ttlb + \De_h) - c \ge 1$. Substituting the FIOS women's IC $\lmr\De_h - c = \ttlb - \lmr\De$ into the latter gives $(1+\lmr\De)(\ttlb - 1) \ge 0$; with $\ttlb < 1$, this forces $1+\lmr\De \le 0$, generically strict.\footnote{The not-fragile conclusion below is likewise unaffected by the correction: it requires only that the fully-investing side's best response stay pinned at the corner $1$ in a neighborhood of the perturbation, which holds because the deviation-optimal IC holds strictly at $\ttlb$ (and hence, by continuity, nearby).} The analogous derivation for NIOS at $(0, \ttub)$ with interior $\ttub > 0$ yields $\ttub(1+\lmr\De) \le 0$, again forcing $1+\lmr\De < 0$.

\emph{SYM AM is fragile.} At the interior fixed point $(\ttsym, \ttsym)$ of $\Theta$ with $\mathrm{BR} = \mathrm{BR}_{AM}$, the symmetric IC gives $\lmr(\De\ttsym + \De_h) - c = \ttsym$, so $\Theta_m(\ttsym + \epsilon, \ttsym - \epsilon) = \mathrm{BR}_{AM}(\ttsym - \epsilon) = \ttsym - \lmr\De\,\epsilon$. Symmetrically, $\Theta_w(\ttsym + \epsilon, \ttsym - \epsilon) = \ttsym + \lmr\De\,\epsilon$. Both deviations have magnitude $\lmr|\De|\,|\epsilon|$. With $\lmr|\De| > 1$ (equivalent under $\De < 0$ to $1+\lmr\De < 0$), Definition~\ref{Definition: fragility} holds with $\zeta = \lmr|\De| - 1 > 0$ and any $\delta$ small enough that the perturbations stay in $(0,1)^2$.

\emph{FIOS is not fragile.} At $(1, \ttlb)$, feasibility requires $\epsilon \le 0$. The men's unclipped best response to $\ttlb - \epsilon$ is
\[
M(\epsilon) \,=\, \lmr(\De(\ttlb - \epsilon) + \De_h) - c \,=\, \ttlb(1+\lmr\De) - \lmr\De - \lmr\De\,\epsilon,
\]
using the FIOS women's IC. At $\epsilon = 0$, $M(0) - 1 = (1+\lmr\De)(\ttlb - 1) = |1+\lmr\De|(1-\ttlb) > 0$. By continuity, $M(\epsilon) > 1$ for $|\epsilon| < \delta_F := |1+\lmr\De|(1-\ttlb)/(\lmr|\De|)$, so $\Theta_m = 1$ throughout this neighborhood. Hence $|\Theta_m(1+\epsilon, \ttlb - \epsilon) - 1| = 0$ for every $\epsilon \in (-\delta_F, 0]$, violating the first inequality of Definition~\ref{Definition: fragility} for any positive $\zeta$. FIOS is not fragile.

\emph{NIOS is not fragile.} The symmetric argument at $(0, \ttub)$, with $\epsilon \ge 0$ the feasible direction. Substituting the NIOS women's IC $\lmr\De_h - c = \ttub$ gives the unclipped men's best response $m(\epsilon) = \ttub(1+\lmr\De) - \lmr\De\,\epsilon$. At $\epsilon = 0$, $m(0) = \ttub(1+\lmr\De) < 0$ with slack $\ttub|1+\lmr\De|$; by continuity $m(\epsilon) < 0$ for $\epsilon \in [0, \delta_N)$ where $\delta_N := \ttub|1+\lmr\De|/(\lmr|\De|)$, so $\Theta_m = 0$ throughout, the first inequality fails, and NIOS is not fragile.
\eprf

\subsection{Derivation of the payoffs in Example~\ref{Example: increasing wages and asymmetric}}\label{Section: example 2 derivation}
This subsection derives the four match payoffs $\phh, \phl, \plh, \pll$ reported in Section~\ref{Section: microfoundation} from the household game. Recall the parameters: $K = 8$, $\a = 0.6$, $t_l = 2$, $t_h \in \{3, 5\}$, and the utility of member $i$ in a match with effort vector $(e_m, e_w)$ is
\begin{align*}
    U_i(e_m, e_w) = K \big[ \a \log( t \cdot e ) + (1-\a) \log(2 - e_m - e_w) \big] - \tfrac{1}{2}(1 - e_i)^2,
\end{align*}
where $t \cdot e = t_m e_m + t_w e_w$ is household income.

\medskip

\noindent \textbf{First-order conditions.} Differentiating $U_i$ with respect to $e_i$,
\begin{align*}
    \frac{\partial U_i}{\partial e_i} = K \a \frac{t_i}{t \cdot e} - K(1-\a)\frac{1}{2 - e_m - e_w} + (1 - e_i).
\end{align*}
A non-cooperative interior equilibrium requires this to vanish for both $i \in \{m, w\}$.

\medskip

\noindent \textbf{Symmetric matches: $(s, s) \in \{(L, L), (H, H)\}$.} When $t_m = t_w =: t$, the model is symmetric in the two members, so $e_m = e_w =: e^*$ and the FOC reduces to
\begin{align*}
    \frac{K \a}{2 e^*} - \frac{K(1-\a)}{2(1 - e^*)} + (1 - e^*) = 0.
\end{align*}
The wage $t$ cancels, so $e^*$ is the same for $(L, L)$ and $(H, H)$. With $K = 8$ and $\a = 0.6$, the equation becomes $\frac{2.4}{e^*} - \frac{1.6}{1 - e^*} + (1 - e^*) = 0$, whose unique root in $(0,1)$ is $e^* \approx 0.6222$. Substituting,
\begin{align*}
    \phi(s, s) &= K \big[\a \log(2 t e^*) + (1-\a) \log(2 - 2 e^*)\big] - \tfrac{1}{2}(1 - e^*)^2 \\
    &= 8 \big[ 0.6 \log(2 t \cdot 0.6222) + 0.4 \log(0.7556) \big] - \tfrac{1}{2}(0.3778)^2.
\end{align*}
Numerically, $\pll = 3.4085$, and $\phh = 5.3547$ at $t_h = 3$ and $\phh = 7.8067$ at $t_h = 5$.

\medskip

\noindent \textbf{Asymmetric matches: $(s_m, s_w) = (H, L)$.} With $t_m = t_h$ and $t_w = t_l$, the wage gap pushes the high-wage member to the boundary $e_m = 1$. The KKT condition for member $m$ at $e_m = 1$ requires $\partial U_m/\partial e_m|_{(1, e_w)} \ge 0$, which at any candidate $(1, e_w)$ reduces to $K\!\left(\a\,\frac{t_h}{t_h + t_l e_w} - (1-\a)\,\frac{1}{1-e_w}\right) \ge 0$; for the parameters considered this holds whenever $e_w \le 0.23$ (at $t_h = 3$) or $e_w \le 0.26$ (at $t_h = 5$)---in particular at the Nash value of $e_w$ identified below ($e_w \approx 0.149$ and $e_w = 0$, respectively), so $e_m = 1$ is member $m$'s best response there.

The Nash structure then depends on member $w$'s FOC at $e_m = 1$.

\noindent \emph{At $t_h = 5$:} the parameters satisfy $\a t_l / t_h < (1 - \a) - 1/K$ (here $\a t_l/t_h = 0.24 < 0.275$), so at $(1, 0)$ member $w$'s marginal payoff $K[\a t_l/t_h - (1-\a)] + 1 = -0.28 < 0$. Member $w$ is also at the boundary, and the Nash equilibrium is the complete corner $(e_m, e_w) = (1, 0)$. Substituting,
\begin{align*}
    \phl &= K \a \log(t_h) - \tfrac{1}{2}(1 - 1)^2 = 4.8 \log(5) = 7.7253, \\
    \plh &= K \a \log(t_h) - \tfrac{1}{2}(1 - 0)^2 = 4.8 \log(5) - 0.5 = 7.2253.
\end{align*}

\noindent \emph{At $t_h = 3$:} the parameters satisfy $\a t_l / t_h > (1 - \a) - 1/K$ (since $\a t_l / t_h = 0.4 > 0.275$), so at $(1, 0)$ member $w$'s marginal payoff $K[\a t_l/t_h - (1-\a)] + 1 = +1 > 0$. Member $w$ strictly prefers to raise $e_w$ above zero, so $(1, 0)$ is \emph{not} Nash. The Nash equilibrium is a partial corner: $e_m = 1$ and $e_w$ solves member $w$'s interior FOC,
\begin{align*}
    K\a \frac{t_l}{t_h + t_l e_w} - K(1-\a) \frac{1}{1 - e_w} + (1 - e_w) = 0.
\end{align*}
Substituting $K = 8$, $\a = 0.6$, $t_h = 3$, $t_l = 2$ and solving numerically yields $e_w \approx 0.1492$. The resulting match payoffs are
\begin{align*}
    \phl &= 8\big[0.6 \log(3 + 2 \cdot 0.1492) + 0.4 \log(1 - 0.1492)\big] \\
         &\quad - \tfrac{1}{2}(1 - 1)^2 \approx 5.2114, \\
    \plh &= 8\big[0.6 \log(3 + 2 \cdot 0.1492) + 0.4 \log(1 - 0.1492)\big] \\
         &\quad - \tfrac{1}{2}(1 - 0.1492)^2 \approx 4.8495.
\end{align*}
The household therefore specializes only partially at $t_h = 3$: the high-wage member supplies all of his time to the market, while the low-wage member supplies a small but positive amount of market labor. At $t_h = 5$, by contrast, complete specialization $(1, 0)$ is the Nash outcome.

\medskip

\noindent \textbf{Verification of the equilibrium claims.} With $r = \lm = 1$ (so $\lmr = 1/2$) and $c = 0.25$:

\noindent For $t_h = 3$: $\De = -1.2977$, $\De_h = 1.8029$, giving
\begin{align*}
    \ttsym = \frac{\lmr \De_h - c}{1 - \lmr \De} = \frac{0.5 \cdot 1.8029 - 0.25}{1 + 0.5 \cdot 1.2977} \approx 0.3951.
\end{align*}
The L-L acceptance constraint at $\ttsym$ is $\lm \ttsym (\plh - \pll) \approx 0.5693 < 3.4085 = r \pll$, so $(\ttsym, \ttsym)$ is the unique equilibrium.

\noindent For $t_h = 5$: $\De = -3.7354$, $\De_h = 4.3168$, giving $\ttsym = 0.6655$ from the same formula. Here $1 + \lmr \De = -0.8677 < 0$, so the FIOS investment IC is feasible; meanwhile the L-L acceptance constraint at $\ttsym$ reads $\lm \ttsym (\plh - \pll) = 0.6655 \cdot 3.8168 = 2.5401 < 3.4085 = r \pll$, so $(\ttsym, \ttsym)$ remains a SYM AM equilibrium. Alongside it, the FIOS cutoff is
\begin{align*}
    \ttlb = \lmr (\De + \De_h) - c = 0.5 \cdot 0.5814 - 0.25 = 0.0407,
\end{align*}
interior and far below $\ttsym$. The investing side's incentive constraint holds, $\min\{\lmr(\ttlb\De+\De_h),\,G(\ttlb)\} = \min\{2.08,\,3.58\} \ge 1 + c$, so the FIOS profile $(1, \ttlb)$ coexists with the SYM AM equilibrium $(\ttsym, \ttsym)$.

\subsection{Proof of Proposition~\ref{Proposition: parametric comparative static}}\label{Section: proof parametric comp static}

\bprf[Proof of Proposition~\ref{Proposition: parametric comparative static}]
Throughout, let $\phi(s, s'; t_h)$ denote the match payoff in the PIH game with skill pair $(s, s')$, where $t_l > 0$ is fixed and $t_h \ge t_l$ varies. We abbreviate $\phh(t_h) = \phi(H, H; t_h)$, $\phl(t_h) = \phi(H, L; t_h)$, $\plh(t_h) = \phi(L, H; t_h)$, and use $\pll = \phi(L, L; t_h)$ which is independent of $t_h$ (the $(L, L)$ match has both members at wage $t_l$, which does not involve $t_h$). Define
\begin{align*}
    \De(t_h)    &:= \phh(t_h) + \pll - \phl(t_h) - \plh(t_h), \\
    \De_h(t_h)  &:= \phl(t_h) - \pll, \\
    \ttsym(t_h) &:= \frac{\lmr \De_h(t_h) - c}{1 - \lmr \De(t_h)}, \\
    \ttlb(t_h)  &:= \lmr [\De(t_h) + \De_h(t_h)] - c, \\
    \ttub(t_h)  &:= \lmr \De_h(t_h) - c.
\end{align*}

\medskip

\noindent \textbf{Step 1: Continuity of the household game's Nash equilibrium.}

Fix any $(t_m, t_w) \in \mathbb{R}_+^2$. By Definition~\ref{Definition: PIH}, the game has a unique Nash equilibrium $e^*(t_m, t_w) \in [0, 1]^2$. Because each $U_i$ is continuous in $(e, t_m, t_w)$ and the strategy space $[0,1]^2$ is compact, the set of Nash equilibria has closed graph in $(t_m, t_w)$; being single-valued by uniqueness, the equilibrium $e^*(\cdot)$ is therefore continuous in $(t_m, t_w)$. Hence $\phi(s, s'; t_h)$ is continuous in $t_h$, and so are $\De$, $\De_h$, $\ttsym$, $\ttlb$, $\ttub$.

\medskip

\noindent \textbf{Step 2: Asymptotic divergence of $\Phi$.}

Define
\begin{align*}
    \Phi(t_h) := \lambda\, \ttsym(t_h)\, (\plh(t_h) - \pll) - r\, \pll.
\end{align*}
The symmetric L-L acceptance \eqref{Equation: AM L accepts L} holds at the symmetric profile iff $\Phi(t_h) \le 0$. We claim $\Phi(t_h) \to +\infty$ as $t_h \to \infty$.

By assumption (C2), $\liminf_{t_h \to \infty} \ttsym(t_h) =: x^* > 0$, so $\ttsym(t_h) \ge x^*/2$ for $t_h$ sufficiently large. Furthermore, $\plh(t_h) \to \infty$: at the household Nash equilibrium $(e_m^*, e_w^*)$ at wages $(t_h, t_l)$, the best-response inequality for member $m$ (using $e_m = 1/2$ as a feasible deviation) gives
\begin{align*}
    U_m(e_m^*, e_w^*;\, t_h, t_l)\ \ge\ \psi\!\left(\tfrac{t_h}{2} + t_l\, e_w^*,\, \tfrac{3}{2} - e_w^*\right) - g(1/2)\ \ge\ \psi(t_h/2,\, 1/2) - g(1/2),
\end{align*}
which tends to $\infty$ by the unboundedness of $\psi$ in its first argument. Since $\plh(t_h) = U_w(e_m^*, e_w^*) = U_m(\cdot) + [g(1-e_m^*) - g(1-e_w^*)]$ and the bracketed cost difference is bounded ($g$ is bounded on $[0,1]$), $\plh(t_h) \to \infty$. Combining with $\ttsym(t_h) \ge x^*/2$ at large $t_h$, $\Phi(t_h) \to \infty$.

Since $\ttsym(t_l) \le 0$ (the unclipped formula gives $-c < 0$) so $\Phi(t_l) \le -r\pll < 0$, and $\Phi$ is continuous (Step 1), the threshold
\[
    t_\Phi := \inf\{t_h \ge t_l : \Phi(t_h) \ge 0\}
\]
is finite. Under the additional hypothesis $\De_h' \ge |\De'|$, $\ttsym(t_h)$ is non-decreasing (Step 5) and $\eta(t_h) = \plh(t_h) - \pll$ is non-decreasing (household income at the $(L,H)$ match rises in $t_h$), so $\Phi(t_h) = \lm\,\ttsym(t_h)\,\eta(t_h) - r\pll$ is non-decreasing and crosses zero exactly once. Thus $t_\Phi$ marks a single AM-to-SA transition of the symmetric equilibrium: for $t_h < t_\Phi$ it is SYM AM, and for $t_h > t_\Phi$ it is SYM SA. (The coexistence claim does not rely on this: a symmetric equilibrium exists at every $t_h$ by Proposition~\ref{Proposition: equilibrium structure}, whether in AM or SA form.)

\medskip

\noindent \textbf{Step 3: Defining $t^*$.}

By (C1), $\De(t_h) \to -\infty$, so $1 + \lmr \De(t_h) \le 0$ for sufficiently large $t_h$. Define $t_h^\dagger := \inf\{t_h \ge t_l : 1 + \lmr \De(t_h) \le 0\}$; this is finite by (C1) and Step 1. By the single-crossing condition (C4), $1 + \lmr\De(t_h) \le 0$ holds for \emph{all} $t_h \ge t_h^\dagger$, not merely at the infimum---so the FIOS investment feasibility, once attained, persists. By (C3), there exists $T_2 < \infty$ such that for all $t_h \ge T_2$, $\ttlb(t_h) \in (0, 1)$ and the FIOS profile $(1, \ttlb(t_h))$ satisfies \eqref{Equation: AM H accepts L} on path. Define
\[
    t^* := \max\{t_h^\dagger,\, T_2\}\ \in (t_l, \infty).
\]
That is, $t^*$ is a threshold above which the three FIOS-existence conditions hold simultaneously: feasibility of the FIOS investment IC ($1 + \lmr \De \le 0$), interior FIOS cutoff ($\ttlb \in (0, 1)$), and FIOS H-L acceptance on path. $t^*$ should not be confused with $t_\Phi$ from Step 2: $t^*$ marks the threshold above which FIOS becomes feasible, while $t_\Phi$ marks the AM-to-SA transition of the symmetric equilibrium. The two may sit on either side of each other in general.

\medskip

\noindent \textbf{Step 4: At $t_h > t^*$, a FIOS equilibrium exists alongside the symmetric one.}

Fix $t_h > t^*$. The FIOS profile is $(1, \ttlb(t_h))$. By the definition of $t^*$, $\ttlb(t_h) \in (0, 1)$, so the cutoff is interior. The women's investment IC at $\ttlb$ is the indifference relation $\lmr [\De + \De_h] = \ttlb + c$, satisfied by definition of $\ttlb$. The men's investment IC at type $1$ is the deviation-optimal condition $\min\{\lmr(\ttlb\De+\De_h),G(\ttlb)\}\ge 1+c$ (the fully-investing side carries no low types on path, so matching is all-match on path and the low type's acceptance is set by optimality; see the definition of the FIOS profile). The all-match term $\lmr[\ttlb\De+\De_h]\ge1+c$ reduces, using $1+\lmr\De\le0$ for all $t_h>t^*$ (Step 3, by (C4)), to $\ttlb\le1$, which holds. For the SA term, $G(\ttlb)=\lmr[\ttlb\phh+(1-\ttlb)\phl]-\lm\ttlb\,\plh/(r+\lm\ttlb)\to\infty$ as $t_h\to\infty$ (the high-type value diverges by the argument of Step 2 while the bracketed coefficient on the divergent payoffs is positive for $\ttlb<1$), so $G(\ttlb)\ge1+c$ for $t_h$ large; raising $T_2$ if necessary, both terms hold for $t_h>t^*$. The H-L acceptance constraint \eqref{Equation: AM H accepts L} is satisfied on path by the definition of $t^*$, and low-skill women meet only high-skill men on path. Hence FIOS is an equilibrium. By Proposition~\ref{Proposition: equilibrium structure} Part 1, a symmetric equilibrium also exists (SYM AM at cutoff $\ttsym$ if $\Phi(t_h) \le 0$, i.e., $t_h \le t_\Phi$; SYM SA at cutoff $\ttsa$ otherwise). The two coexist.

\medskip

\noindent \textbf{Step 5: Within-regime comparative statics.}

\noindent \emph{Monotonicity of $\ttsym(t_h)$ in the symmetric AM regime $[t_l, t_\Phi)$ (under the additional hypothesis $\De_h' \ge |\De'|$).} On this regime, the cutoff is given by the closed form
\begin{align*}
    \ttsym(t_h) = \frac{\lmr \De_h(t_h) - c}{1 - \lmr \De(t_h)} = \frac{N(t_h)}{D(t_h)},
\end{align*}
where $N(t_h) := \lmr \De_h(t_h) - c$ and $D(t_h) := 1 - \lmr \De(t_h)$. Under the smoothness of the PIH game (Step 1), both $N$ and $D$ are continuously differentiable in $t_h$. By the quotient rule,
\begin{align*}
    \frac{d \ttsym}{d t_h} = \frac{N'(t_h)\, D(t_h) - N(t_h)\, D'(t_h)}{D(t_h)^2} = \frac{\lmr \big[\De_h'(t_h)\, D(t_h) + \De'(t_h)\, N(t_h)\big]}{D(t_h)^2}.
\end{align*}
Within the symmetric regime, $\ttsym(t_h) = N(t_h)/D(t_h) \in [0, 1)$, which (combined with $D > 0$) gives $0 \le N(t_h) \le D(t_h)$. Under the hypothesis $\De_h'(t_h) \ge |\De'(t_h)|$ (for the Cobb-Douglas case this holds with equality at the full corner $t_h \gtrsim 4.4$, where $\De_h'(t_h) = -\De'(t_h) = K\a/t_h$, and strictly for smaller $t_h$), and using $\De'(t_h) \ge -|\De'(t_h)|$,
\begin{align*}
    \De_h'(t_h)\, D(t_h) + \De'(t_h)\, N(t_h)
    &\ge \De_h'(t_h)\, D(t_h) - |\De'(t_h)|\, N(t_h) \\
    &\ge |\De'(t_h)|\, D(t_h) - |\De'(t_h)|\, N(t_h) \\
    &= |\De'(t_h)|\, [D(t_h) - N(t_h)] \ \ge\ 0.
\end{align*}
Hence $d\ttsym/dt_h \ge 0$, and $\ttsym$ is non-decreasing in $t_h$ on the AM regime $[t_l, t_\Phi)$.

\medskip

\noindent \emph{The FIOS cutoff lies below the symmetric one.} The SYM AM indifference condition $\lmr(\De\,\ttsym + \De_h) - c = \ttsym$ and the FIOS cutoff $\ttlb = \lmr(\De + \De_h) - c$ together give $\ttsym - \ttlb = \lmr\De\,(\ttsym - 1) = \lmr|\De|\,(1 - \ttsym)$, strictly positive whenever $\De < 0$ and $\ttsym < 1$. Hence, wherever a FIOS equilibrium exists alongside SYM AM, its cutoff lies strictly below the symmetric cutoff. Finally $\ttlb(t_h) = \lmr[\De(t_h) + \De_h(t_h)] - c$ is non-decreasing in $t_h$, since $d\ttlb/dt_h = \lmr(\De_h' + \De') \ge \lmr(\De_h' - |\De'|) \ge 0$ under the slope hypothesis.

\medskip

\noindent \emph{Discontinuous appearance of FIOS at $t_h = t^*$.} The FIOS cutoff is $\ttlb(t_h) = \lmr [\De(t_h) + \De_h(t_h)] - c = \lmr[\phh(t_h) - \plh(t_h)] - c$, which stays small as the premium rises---in the Cobb-Douglas case $\phh - \plh$ becomes independent of $t_h$ once the $(H,L)$ household fully specializes ($t_h \gtrsim 4.4$), where the $\log t_h$ terms cancel, so $\ttlb$ levels off near $0.04$. At $t_h = t^* \approx 3.51$, the symmetric cutoff is $\ttsym(t^*) \approx 0.51$, far above $\ttlb \approx 0.03$. The discontinuity is not in any single equilibrium cutoff (the symmetric trajectory is continuous, as shown above) but in the equilibrium \emph{set}: a new FIOS equilibrium emerges at $t^*$, far from the symmetric one. If the population coordinates on the FIOS equilibrium for $t_h$ slightly above $t^*$, the investment profile jumps discretely from $(\ttsym, \ttsym) \approx (0.51, 0.51)$ to $(1, \ttlb) \approx (1, 0.03)$: one gender goes from $\ttsym$ to full investment, the other drops from $\ttsym$ to $\ttlb \ll \ttsym$.

\medskip

Steps 3--4 establish that a FIOS equilibrium coexists with the symmetric equilibrium for all $t_h > t^*$; and since both asymmetric forms require $1 + \lmr\De < 0$ (the FIOS and NIOS investment ICs), no asymmetric equilibrium exists while $1 + \lmr\De(t_h) > 0$. Step 5 establishes the comparative statics on the SYM AM range: the symmetric cutoff $\ttsym$ is continuous and non-decreasing in $t_h$, the FIOS cutoff satisfies $\ttlb < \ttsym$ wherever it exists, and $\ttlb(t_h) = \lmr(\De + \De_h) - c$ is itself non-decreasing (since $d\ttlb/dt_h = \lmr (\De_h' + \De') \ge 0$ whenever $\De_h' \ge |\De'|$). The threshold $t^*$ satisfies $t_l < t^* < \infty$.
\eprf

\newpage
\begin{center}
\rule{0.7\textwidth}{0.4pt}\\[2pt]
{\Large\bfseries Online Appendix}\\[2pt]
\rule{0.7\textwidth}{0.4pt}
\end{center}
\medskip

\section{PAM equilibrium}\label{Section: PAM and SA appendix}

This section presents the formal analysis of the PAM (positive assortative matching) equilibrium class, referenced from the Remark following Proposition~\ref{Proposition: equilibrium structure}.

\subsection*{Setup}\label{Section: PAM analysis}

A PAM equilibrium is one in which high-skilled agents reject low-skilled partners ($\accept^\g(H) = 0$) while low-skilled agents accept low-skilled partners ($\accept^\g(L) = 1$); since the high types reject them, low types are matched only with low types. The players' payoffs (with some abuse of notation), denoted by $\valfn^\g_P(\cdot)$, are:
\begin{align}
    \valfn_P^\g(H; \ttype) &= \frac{\lm \ttypec_{\gt} \phh}{r + \lm \ttypec_{\gt}}, \tag{PAM: Values} \label{Equation: PAM Values} \\
    \valfn_P^\g(L; \ttype) &= \frac{\lm (1-\ttypec_{\gt}) \pll}{r + \lm (1-\ttypec_{\gt})}.\notag
\end{align}
For such an equilibrium to exist with an interior $\ttype$, the IC constraints are:
\begin{align}
    \valfn^\g_P(H; \ttype) - \cost(\ttypecg) =&\ \valfn^\g_P(L; \ttype) \tag{PAM: Invest} \label{Equation: PAM invest IC}\\
    \valfn^\g_P(H; \ttype) \ \ge&\ \phl \tag{PAM: Assortativity}\label{Equation: PAM assortativity}
\end{align}
Here, the investment IC checks for the indifference of the cutoff type $\ttypecg$, while the assortativity constraint ensures that a high-skill person prefers to wait for another high-skill partner than to match with a low-skill one.

\bprop\label{Proposition: PAM is symmetric} No PAM exists if $\lm [ \phh - \phl ] < r \phl$. Whenever a PAM exists, it is symmetric, i.e., $\ttypem = \ttypew$. \eprop

\bprf[Proof of Proposition~\ref{Proposition: PAM is symmetric}]
From \eqref{Equation: PAM assortativity}, we have that
\begin{align*}
    \frac{\lm \ttypec_{\gt} \phh}{r + \lm \ttypec_{\gt}} \ge \phl
\end{align*}
The LHS is increasing in $\ttypec_{\gt}$, and hence, is maximized at $1$. Substituting $\ttypec_{\gt} = 1$, we get that $\valfn^\g(H;\ttype)$ is maximized at $\frac{\lm \phh}{r + \lm}$. Thus, if $\lm [ \phh - \phl ] < r \phl$, then \eqref{Equation: PAM assortativity} cannot be satisfied for any $\ttype$ establishing the first part of the Proposition.

For the symmetry of a PAM whenever it exists, suppose $\ttype = (\ttypem, \ttypew)$ is a PAM equilibrium with $\ttypem \ne \ttypew$. Without loss of generality, $\ttypem > \ttypew$, which forces $\ttypem > 0$ and $\ttypew < 1$. The investment IC \eqref{Equation: PAM invest IC} at the cutoffs (in its boundary-inclusive form) gives,
\begin{align*}
    \valfn^m_P(H; \ttype) - \valfn^m_P(L; \ttype) &\ge \cost(\ttypem) \quad \text{(type } \ttypem > 0 \text{ invests)},  \\
    \valfn^w_P(H; \ttype) - \valfn^w_P(L; \ttype) &\le \cost(\ttypew) \quad \text{(type just above } \ttypew < 1 \text{ does not)}.
\end{align*}
Subtracting the second from the first,
\begin{align*}
    [\valfn^m_P(H; \ttype) -  \valfn^m_P(L; \ttype)] - [\valfn^w_P(H; \ttype) -  \valfn^w_P(L; \ttype)] &\ge \cost(\ttypem)-\cost(\ttypew) > 0,
\end{align*}
where the final inequality uses $\cost$ strictly increasing and $\ttypem > \ttypew$. On the other hand, from \eqref{Equation: PAM Values},
\begin{align*}
&\left[\valfn^m_P(H; \ttype)  - \valfn^m_P(L; \ttype)   \right] - \left[\valfn^w_P(H; \ttype)  - \valfn^w_P(L; \ttype)   \right] \\
&=\left[\frac{\lm \ttypew\phh}{r + \lm \ttypew} - \frac{\lm (1- \ttypew) \pll}{r + \lm (1-\ttypew)}\right] \\
&\quad - \left[\frac{\lm \ttypem\phh}{r + \lm \ttypem} - \frac{\lm (1- \ttypem) \pll}{r + \lm (1-\ttypem)}\right] = h(\ttypew) - h(\ttypem),
\end{align*}
where $h(x) := \frac{\lm x\phh}{r + \lm x} - \frac{\lm (1- x) \pll}{r + \lm (1-x)}$ is strictly increasing. Since $\ttypem > \ttypew$, $h(\ttypew) - h(\ttypem) < 0$, contradicting the previous inequality.
\eprf

\end{document}